\documentclass[12pt]{scrartcl}
\usepackage{pslatex}
\usepackage{cite}
\usepackage{graphicx}
\usepackage{amsmath}
\allowdisplaybreaks[1] 
\graphicspath{{./figs/}}

\def\as{\alpha_s}
\def\Eq#1{Eq.~(\ref{#1})}
\def\Refs#1{Refs.~\cite{#1}}
\def\Ref#1{Ref.~\cite{#1}}
\def\Fig#1{Fig.~\ref{#1}}

\def\Tab#1{Tab.~\ref{#1}}

\def\kt{{k_t}}
\def\kb{{k_b}}
\def\ku{{k_u}}
\def\kd{{k_d}}
\def\ubar{\overline{u}}
\def\slashed#1{{\,/\!\!\!\!\!#1}}
\def\S#1{{\cal S}_{#1}}
\def\Shat#1{{\hat{S}}_{#1}}
\def\Tr{\mbox{Tr}}

\def\LO{{\mbox{\scriptsize LO}}}
\def\NLO{{\mbox{\scriptsize NLO}}}

\def\mt{m_t}
\def\mw{m_W}
\def\mz{m_Z}

\begin{document}
\begin{flushright}
  SFB/CPP-14-71\\
  HU-EP-14/32
\end{flushright}
\begin{center}
  {\Large\bfseries
    Calculation of two-loop QCD corrections for hadronic\\[0.2cm] 
    single top-quark production in the $t$ channel}
  \vspace*{0.3cm}

  M.~Assadsolimani, P.~Kant, B.~Tausk, P.~Uwer
  \vspace*{0.1cm}

  Humboldt-Universit\"at zu Berlin, Institut f\"ur Physik,\\
  Newtonstra\ss e 15, D-12489 Berlin, Germany

  \vspace*{0.2cm}

  \begin{center}
    {\bfseries Abstract:}\\[1em]
    \parbox{10cm}{
In this article we discuss the calculation of single top-quark
production in the $t$ channel at two-loop order in QCD. In particular
we present the decomposition of the amplitude according to its spin and
colour structure and present complete results for the two-loop
amplitudes in terms of master integrals. For the vertex corrections
compact analytic expressions are given. The box contributions are
implemented in a publicly available C program.
      }
  \end{center}
\end{center}
\section{Introduction}
In hadronic collisions top quarks are dominantly produced in pairs
through the flavour conserving strong interaction. The charged
currents of the weak interaction allow to produce top-quarks or
anti-quarks also singly.  Since the threshold for single top-quark
production is half that of top-quark pair production, the weak
coupling is partially compensated by phase space effects, and the
larger parton fluxes such, that for LHC operating at 13 TeV the cross
section for single top-quark production is roughly one third of the
top-quark pair cross section.  Despite these significant event rates, single
top-quark production is experimentally challenging due to the
complicated experimental signature and the sizeable backgrounds. To
cope with these difficulties sophisticated experimental techniques
like for example the matrix element method and methods making use of
neural networks have been developed at the Tevatron and at the LHC
\cite{Dalitz:1991wa,Dalitz:1992np,Kondo:1993in,Abazov:2004cs}.
Using these techniques, very precise measurements will be possible at
the LHC running at 13 TeV. Physics wise single top-quark production is highly
interesting since it allows a precise test of the top-quark weak
couplings. In particular, single top-quark production offers a unique
source of highly polarized top quarks. Furthermore the
Cabibbo-Kobayashi-Maskawa (CKM) matrix element $V_{tb}$ is directly
accessible without further assumptions. In addition, single top-quark
production can be used to constrain the bottom quark distribution inside
the proton. 

Depending on the momentum of the $W$ boson involved in the charged
current interaction, three different contributions to single top-quark
production can be distinguished: the $t$-channel contribution for
space-like momentum, the $s$-channel contribution for time-like
momentum and the $tW$ channel for on-shell production of a $W$ boson
in association with the top quark. At both colliders, Tevatron and
LHC, the $t$-channel gives the largest contribution to the cross
section. The second important channel at the LHC is $tW$ production.
This contribution is suppressed at the Tevatron due to the limited
energy of the collider. The $s$-channel contributes about 30-40\% to
the cross section at the Tevatron while it is small at the LHC. 
\begin{table}[htbp]
  \begin{center}
    \leavevmode\renewcommand{\arraystretch}{1.4}
 \begin{tabular}[htbp]{l|c|c|c|c|c|c}
      &\multicolumn{4}{c|}{LHC 13 TeV} &\multicolumn{2}{c}{Tevatron} \\ 
      & $\sigma_t^\LO$ & $\sigma_{\bar t}^\LO$ & 
      $\sigma_t^\NLO$ & $\sigma_{\bar t}^\NLO$ & $\sigma_{t,\bar t}^\LO$ &
      $\sigma_{t,\bar t}^\NLO$ \\ \hline 
      $t$ 
      &135 &79.8 &137\,$^{+4.0}_{-2.3}$ $^{+1.0}_{-0.9}$ &
      82.1\,$^{+2.5}_{-1.3}$ $^{+0.6}_{-0.8}$ 
      &$1.03$
      & 0.998\,$^{+0.025}_{-0.022}$ $^{+0.029}_{-0.032}$\\ \hline
      $s$ &4.27 &2.63 &
      6.25\,$^{-0.06}_{+0.09}$ $^{+0.12}_{-0.09}$
      &3.97\,$^{-0.04}_{+0.05}$ $^{+0.08}_{-0.07}$
        &0.28 &0.442\,$^{-0.023}_{+0.025}$ $^{+0.015}_{-0.011}$\\ \hline
      $tW$  &29.1 &29.1 &29.3\,$^{+1.0}_{-1.3}$ $^{+0.7}_{-0.8}$
      &29.2\,$^{+1.0}_{-1.3}$ $^{+0.7}_{-0.8}$
      &0.069& 0.070\,$^{-0.002}_{-0.001}$ 
      $^{+0.008}_{-0.009}$\\ \hline
    \end{tabular}
    \caption{Cross sections for single top-quark production in pb,
      for $m_t=173.3$~GeV, $\mu_R=\mu_f=m_t$ and the MSTW2008lo/nlo
      PDF set, obtained using the Hathor program \cite{Kant:2014oha}.
      The sub- and superscripts denote the uncertainty due to scale
      variation and PDF uncertainties.}
    \label{tab:xsvalues}
  \end{center}
\end{table}
The situation is summarized in \Tab{tab:xsvalues} where also
predictions at next-to-leading order (NLO) accuracy are given. The NLO
corrections have been calculated for the different channels in
\Refs{Bordes:1994ki,Smith:1996ij,Stelzer:1997ns,Stelzer:1998ni,%
Harris:2002md,Sullivan:2004ie,Sullivan:2005ar,Giele:1995kr,Zhu:2002uj}.
The corrections to the $t$-channel are very small. However,
the small size of the corrections is due to a significant
cancellation between individual contributions. Furthermore, only the
vertex corrections contribute to the cross section. The box-type
corrections to the amplitude vanish when interfered with the Born
amplitude. In particular, no colour exchange between the two incoming
quark lines is possible in the $t$-channel. As a consequence it is
conceivable that the small size of the NLO corrections is accidental
and significant contributions at NNLO could appear when colour exchange
between the two quark lines becomes possible. In particular,
differential distributions which are crucial to test the $V-A$
structure of the weak interaction and the polarization of the
top-quark imprinted by the production mechanism, may be affected
significantly once the NNLO QCD corrections are taken into account.
Very recently partial results for single top-quark production at NNLO
accuracy have been presented for the $t$-channel in
\Ref{Brucherseifer:2014ama}. The analysis is restricted to the vertex
corrections and the related real corrections.  The corrections to the
inclusive cross sections are at the level of one per cent and thus
small from a phenomenological point of view. However, compared to the
NLO corrections, they amount to about 50\% and are thus much larger
than one would naively expect --- indicating that the full NNLO
corrections may indeed give important corrections. The full NNLO
corrections require the evaluation of a variety of different
contributions: two-loop amplitudes interfered with the Born amplitude,
corrections due to one-loop amplitudes squared, one-loop corrections
to the real emission processes, and finally double real emission
processes.  For most of the contributions established methods for
their evaluation exist. In some cases even public tools are available to
perform the required calculations. As far as the two-loop amplitudes
are concerned the situation is more involved. In principle, techniques
exist for the reduction of two-loop tensor integrals to a small set of
master integrals.  However, in practice this reduction and the
evaluation of the master integrals is highly non-trivial and has been
solved in the past only on a case by case basis. Depending on the
number of invariants and masses the complexity of the reduction
increases. In this article we present the complete reduction of the
two-loop amplitudes for the $t$-channel. Results for the $s$-channel
can be obtained through crossing or by adapting the reduction
presented here. In section 2 we discuss the decomposition of the
two-loop amplitudes according to the spin and colour structure. We also
comment on our treatment of $\gamma_5$ in $d$ space-time dimensions.
In section 3 we present some details about the integral reduction. In
section 4 we present the complete 2-loop amplitude in terms of
master integrals further decomposed according to colour and spin
structure. For the  vertex corrections compact analytic results
are given. As far as the double box contributions are
concerned we briefly discuss their numerical evaluation using an
implementation in C which can be obtained on demand. Due to their length
it is not useful to present analytic expressions here. We close with a
short conclusion in section 5.

\section{Theoretical setup}
\begin{figure}[htbp]
  \begin{center}
    \includegraphics{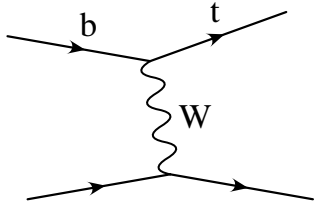}
    \caption{Born diagram for $t$-channel production of a single top quark.}
    \label{fig:borndiagram}
  \end{center}
\end{figure}
In this section we briefly describe the theoretical setup used to
organise the calculation of the two-loop amplitudes for
single top-quark production in the $t$-channel.
To fix our conventions, we consider the partonic process
\begin{equation}
\label{eq:partonicprocess}
  u(\ku) + b(\kb) \to t(\kt) + d(\kd) \, , 
\end{equation}
treating all quarks --- with the exception of the top
quark --- as massless. We define two Mandelstam variables by
\begin{equation}
  \hat t = (\kb-\kt)^2 = \mt^2-2\kb\kt,\quad  
  \hat s = (\ku+\kb)^2 = 2\ku\kb \, .
\end{equation}
At leading order in the electroweak coupling $g_W$, the amplitude
for this process is expanded in the strong coupling constant
$\as={g_s^2\over 4\pi}$:
\begin{equation}
{\cal A} = g_W^2 V_{tb} V^{*}_{ud}
\left( {\cal A}^{(0)}
      + \frac{\alpha_s}{4\pi} {\cal A}^{(1)}
      + \left(\frac{\alpha_s}{4\pi}\right)^2 {\cal A}^{(2)}
      + \ldots \right) .
\end{equation}
The Born contribution, shown in \Fig{fig:borndiagram}, is given by
\begin{equation}
\label{eq:0loop-color}
{\cal A}^{(0)} = \delta_{tb} \delta_{du} A_{1}^{(0)} ,
\end{equation}
with
\begin{equation}
A_{1}^{(0)} =
\frac{1}{\hat t-\mw^2} \bar u(\kt) \gamma_\mu \frac{1}{2}(1-\gamma_5) u(\kb)
  \bar u(\kd) \gamma^\mu \frac{1}{2}(1-\gamma_5) u(\ku) \, .
\end{equation}
The electroweak coupling can be
expressed in terms of the electric charge $e$ of a positron and the sine of the
Weinberg mixing angle $\vartheta_W$ through
\begin{equation}
  g_W = {e\over \sqrt{2}\sin(\vartheta_W)}.
\end{equation}
Working in leading order in the electroweak coupling, the
renormalization scheme of the electroweak parameters is not fixed. For
phenomenological applications one may use the on-shell scheme in which
the weak mixing angle can be calculated from
the mass of the $Z$-boson ($\mz$) and the mass of the $W$-boson
($\mw$) using: 
\begin{equation}
  \cos^2(\vartheta_W) = {\mw^2\over \mz^2}. 
\end{equation}
The matrix elements of the Cabibbo-Kobayashi-Maskawa matrix, which
expresses the eigenstates of the weak interaction in terms of the
mass eigenstates, are denoted by $V_{ij}$.
Since the Born amplitude is a purely electroweak process, no colour
exchange between the two quark lines is possible. This is reflected in
the colour structure $\delta_{tb}\delta_{du}$ where $t,b,\ldots$
describe the colour indices of the respective quarks and $\delta$ denotes the
Kronecker delta. However, when higher order QCD corrections are included,
colour exchange between the two quark lines does become possible.
\begin{figure}[htbp]
  \begin{center}
  \includegraphics{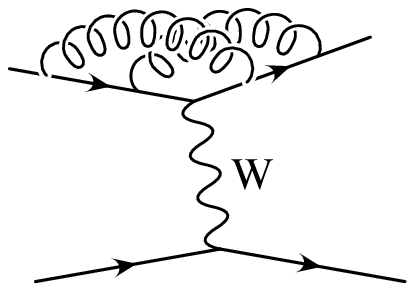}\hspace*{1.2cm}
  \includegraphics{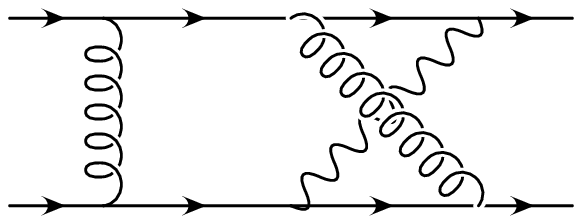}
  \vspace*{0.8cm}

  \includegraphics{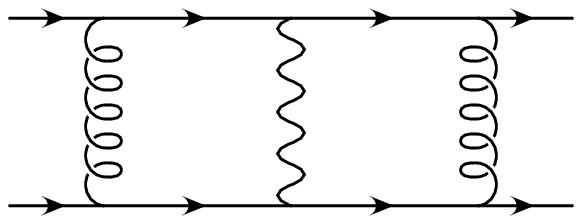}\hspace*{0.3cm}
  \includegraphics{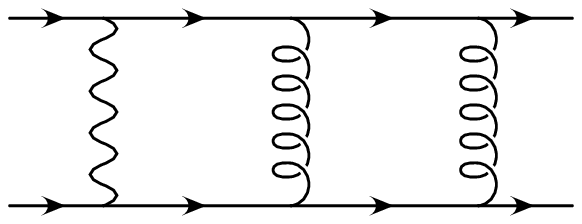}
  \caption{Sample diagrams for single top-quark production at two-loop order.}
  \label{fig:twoloop}
\end{center}
\end{figure}
\begin{figure}[htbp]
  \begin{center}
  \hspace*{-4cm}
  \includegraphics{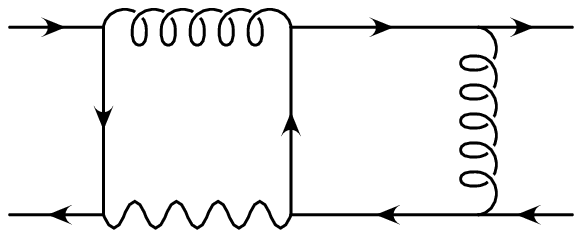}\hspace*{0.3cm}
  \includegraphics{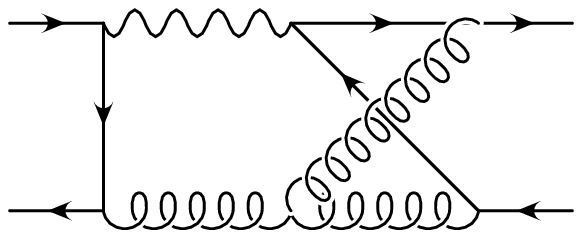}\hspace*{0.3cm}
  \begin{picture}(0,0)
    \put(0,-20){\includegraphics{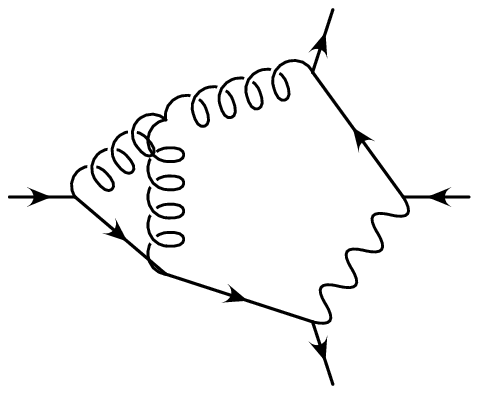}}
  \end{picture}
  \vspace*{0.3cm}
  \caption{Diagrams whose interference with the Born amplitude vanishes.}
  \label{fig:vanishingtwoloop}
\end{center}
\end{figure}

\begin{figure}[htbp]
  \begin{center}
    \leavevmode
    \includegraphics[scale=0.9]{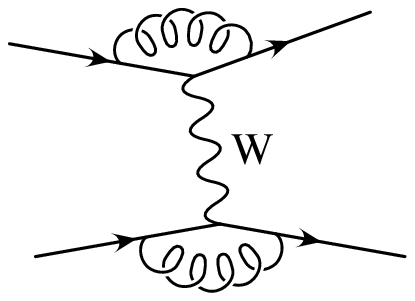}
    \includegraphics[scale=0.9]{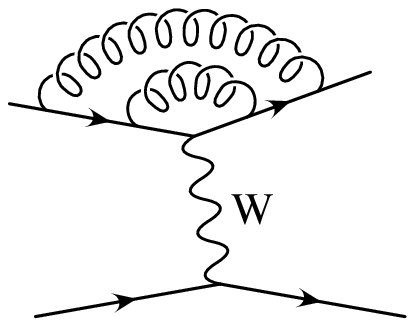}
    \includegraphics[scale=0.9]{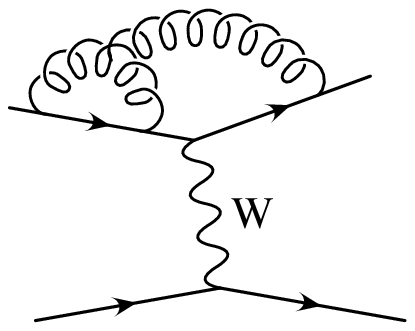}
    \includegraphics[scale=0.9]{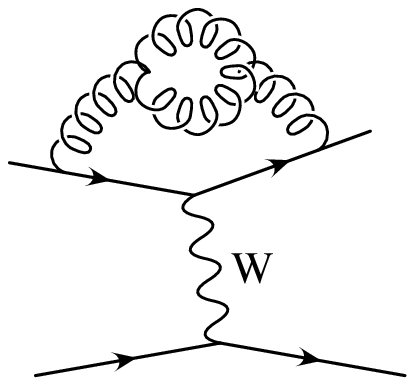}
    \caption{Sample topologies contributing to $A_{1,LC}^{(2)}$.}
    \label{fig:LC-topos}
  \end{center}
\end{figure}

In two-loop approximation about 70 Feynman diagrams contribute to the
transition matrix element --- not counting self-energy corrections and
counter term diagrams. Sample diagrams are shown in
\Fig{fig:twoloop} and \Fig{fig:vanishingtwoloop}.
As we shall see in the following, many diagrams do not contribute to
the cross section at NNLO accuracy.
The colour decomposition of the two-loop amplitude reads:
\begin{equation}
  \label{eq:2loop-color}
  {\cal A}^{(2)} = \delta_{tb}\delta_{du}\, A_1^{(2)} +
  \left(\delta_{tu}\delta_{db} - {1\over
      N}\delta_{tb}\delta_{du}\right)
  A_2^{(2)},
\end{equation}
where $N$ denotes the number of colours. We work with the
general SU(N) gauge group, to make the colour
structure more explicit. The QCD case is obtained by setting $N=3$. 
Using the Fierz identity 
\begin{equation}
  (T^a)_{ij} (T^a)_{kl} = T_r \left(\delta_{il}\delta_{kj}
  - {1\over N}\delta_{ij}\delta_{kl} \right),
\end{equation}
where $T^a$ ($a=1\ldots N^2-1$) are the generators of SU(N) in the
fundamental representation (we use the normalization
$\mbox{tr}(T^a T^b)=T_r \delta_{ab}$, with $T_r={1\over 2}$), 
it follows that all Feynman diagrams
where only one gluon is exchanged between the two quark lines
contribute only to $A_2^{(2)}$. Because the interference of
the $A_2^{(2)}$-term with the colour structure of the Born
amplitude vanishes, this means that such diagrams do not
contribute to the cross section at order $\as^2$. Some
examples are shown in \Fig{fig:vanishingtwoloop}.
This is similar to the one-loop corrections, where
box-topologies do not contribute to the cross section in order $\as$.
(Note however, that the one-loop box diagrams squared start to
contribute to the cross section at order $\as^2$.)  Using the colour
decomposition as shown in \Eq{eq:0loop-color} and \Eq{eq:2loop-color}
we obtain for the interference of the two-loop amplitude with the Born
amplitude:
\begin{equation}
  \sum_{\mbox{\scriptsize color}} {{\cal A}^{(0)}}^\ast 
  {\cal A}^{(2)} = N^2 {A^{(0)}_1}^\ast A^{(2)}_1.
\end{equation}
Making the $N$ dependence of $A^{(2)}_1$ explicit it is possible to
decompose $A_1^{(2)}$ further into leading- and sub-leading colour
contributions:
\begin{equation}
\label{eq:ColorDecomposition}
  A_1^{(2)} = (N^2-1) \left( A^{(2)}_{1,LC}  
    + {1\over N} T_r (B_h + n_l B_l) + {1\over N^2} A^{(2)}_{1,SC}\right)
\end{equation}
The contributions $B_h$, $B_l$ are due to self-energy insertions in the
one-loop topologies. $B_l$ is due to massless quarks ($n_l$ counts the
number of massless quark flavours), while $B_h$ is due to a top-quark loop in
the gluon propagator.  Interfering the full two-loop amplitude with
the Born amplitude we thus obtain
\begin{equation}
  \sum_{\mbox{\scriptsize color}} {{\cal A}^{(0)}}^\ast 
  {\cal A}^{(2)} = N^2 (N^2-1) \left({A^{(0)}_1}^\ast A^{(2)}_{1,LC}
    + {T_r\over N}  {A^{(0)}_1}^\ast  (B_h + n_l B_l)
    + {1\over N^2}{A^{(0)}_1}^\ast A^{(2)}_{1,SC}\right).
\end{equation}
The surviving diagrams can be classified into
the following different groups:
\begin{enumerate}
\item Gluonic self-energy corrections due to closed quark loops 
  inserted into one-loop topologies, contributing to $B_h$ and $B_l$,
\item  A single diagram, consisting of two one-loop vertex corrections
  to the two weak vertices (first diagram in \Fig{fig:LC-topos}), 
  contributing to $A^{(2)}_{1,LC}$ and $A^{(2)}_{1,SC}$,
\item Two-loop vertex corrections,
  contributing to $A^{(2)}_{1,LC}$ and $A^{(2)}_{1,SC}$,
\item Planar and non-planar double-box topologies,
  contributing to $A^{(2)}_{1,SC}$.
\end{enumerate}
In \Fig{fig:LC-topos} sample topologies contributing to $A_{1,LC}$ are
shown. The Feynman diagrams involving the three gluon vertex
contribute only to $A^{(2)}_{1,LC}$ while the other topologies
contribute to $A^{(2)}_{1,LC}$ and $A^{(2)}_{1,SC}$.

Let us now discuss the spin structure of the amplitude, which we compute
in Feynman gauge, using dimensional regularization to regularize ultraviolet
and infrared divergences. A typical term in the calculation consists of
spinors and gamma matrices, contracted with tensor integrals of the form
\begin{align}
\label{eq:tensorintegraldef}
& I_n(d,\nu_1,\ldots,\nu_n)[
\ell_1^{\alpha_1}\cdots\ell_1^{\alpha_{N_1}}
\ell_2^{\beta_1} \cdots\ell_2^{\beta_{N_2}}
]
=
\int d^{d}\ell_1 d^{d}\ell_2
\frac{
\ell_1^{\alpha_1}\cdots\ell_1^{\alpha_{N_1}}
\ell_2^{\beta_1} \cdots\ell_2^{\beta_{N_2}} }
{ {D_1}^{\nu_1} \cdots {D_n}^{\nu_n} } \; ,
\end{align}
with scalar propagator denominators $D_i$ which are functions of the
loop momenta $\ell_1,\ell_2$ and of the external momenta $k_j$.
The propagators appear with powers $\nu_i$ which may be different from one. 
The tensor integrals in eq.~(\ref{eq:tensorintegraldef}) can be reduced to
a basis of tensors constructed from metric tensors and the external
momenta $k_j$
\begin{align}
\label{eq:basictensor}
I_n(d,\nu_1,\ldots,\nu_n)[
\ell_1^{\alpha_1}\cdots\ell_1^{\alpha_{N_1}}
\ell_2^{\beta_1} \cdots\ell_2^{\beta_{N_2}}
 ] & = \sum_{i} C_i
T_i^{
\alpha_1 \cdots \alpha_{N_1}
\beta_1  \cdots \beta_{N_2}
} \, ,
\end{align}
with scalar coefficients $C_i$.
Next, the basic tensors
$ T_i^{
\alpha_1 \cdots \alpha_{N_1}
\beta_1  \cdots \beta_{N_2} } $
are contracted with the spinors and gamma matrices, and the resulting
expressions are simplified by applying rules of the Dirac algebra.

At this point we need to specify how $\gamma_5$ is treated
in $d$-dimensional space-time. It is well known that $\gamma_5$ is an
intrinsically four-dimensional object and cannot be generalized in a
smooth way to $d$ dimensions.
Different prescriptions to deal with
$\gamma_5$ can be found in the literature (see for example
\Refs{tHooft:1972fi,Breitenlohner:1977hr,Chanowitz:1979zu,Bonneau:1980yb,%
Kreimer:1989ke,Korner:1991sx}).
In the 't~Hooft-Veltman scheme or variants thereof an explicit
definition of $\gamma_5$ in $d$ dimensions is given. For example in the
't~Hooft-Veltman scheme one uses
\begin{equation}
  \gamma_5 = i \gamma_0\gamma_1\gamma_2\gamma_3.
\end{equation}
In $d$ dimensions we now have
\begin{equation}
  \{\gamma_5,\gamma_\mu\} = \gamma_5\gamma_\mu+\gamma_\mu\gamma_5 = 0 
  \mbox{ for } \mu=0,\ldots,3
\end{equation}
and
\begin{equation}
  [\gamma_5,\gamma_\mu] = 0, \mbox{ for } \mu \neq 0,\ldots,3 .
\end{equation}
In general, the naive application of such a scheme violates Ward
identities involving the axial-vector current, since the derivation of
the same typically relies on formal manipulations using an
anti-commuting $\gamma_5$. These Ward identities have to be restored
using additional counter terms (see for example \Ref{Larin:1993tq}) unless
they correspond to observable effects related to axial-vector current
anomalies.  In contrast, a prescription which ensures
$\{\gamma_5,\gamma_\mu\} = 0$, guarantees that axial-vector Ward
identities are reproduced. However, using anti-commutativity together
with the cyclicity of the trace operation it is trivial to show that
\begin{equation}
  \Tr(\gamma_5 \gamma_\alpha\gamma_\beta\gamma_\gamma\gamma_\delta)=0
\end{equation}
for $d\not=4$. A way out is to choose the scheme as proposed in
\Refs{Kreimer:1989ke,Korner:1991sx} where the cyclicity of the trace
is given up. In fact as pointed out in \Ref{Veltman:1988au} 
the method proposed in
\Ref{Chanowitz:1979zu} can also be understood as a modification of the
trace operation. An anti-commuting $\gamma_5$ obviously fails to
reproduce the Adler-Bell-Jackiw anomaly. Since in the present calculation no
anomalous contribution arises we use $\{\gamma_5,\gamma_\mu\}= 0$.

Using an anti-commuting $\gamma_5$ in $d$ dimensions, and taking into
account that only terms with an odd number of Dirac matrices along the
massless quark line connecting the external $d$ and $u$ quarks can appear,
we find that the spinor structures in any of the two-loop diagrams can
be reduced to a linear combination of the following 11 basic structures:
\begin{eqnarray}
  \S{1} &=& \ubar(\kt)\,\gamma_{7}\,u(\kb)\times
  \ubar(\kd)\, \gamma_{6} \slashed{\kt}\,u(\ku)\nonumber \\
  \S{2} &=& \ubar(\kt)\, \gamma_{6} \slashed{\ku} \,u(\kb)\times
  \ubar(\kd)\,\gamma_{6} \slashed{\kt} \,u(\ku)\nonumber \\
  \S{3} &=&\ubar(\kt)\, \gamma_{6} \gamma_{\mu_1} \,u(\kb)\times
  \ubar(\kd)\,\gamma_{6} \gamma_{\mu_1} \,u(\ku)\nonumber \\
  \S{4} &=& \ubar(\kt)\, \gamma_{7}
  \gamma_{\mu_{1}}\slashed{\ku}\,u(\kb)\times
  \ubar(\kd)\, \gamma_{6} \gamma_{\mu_{1}} \,u(\ku)\nonumber \\
  \S5&=&\ubar(\kt)\, \gamma_{7} \gamma_{\mu_{1}}\gamma_{\mu_{2}}\,u(\kb)\times
  \ubar(\kd)\, \gamma_{6}\gamma_{\mu_{1}}\gamma_{\mu_{2}}\slashed{\kt} \,u(\ku)
  \nonumber \\
  \S6&=&\ubar(\kt)\, \gamma_{6} \gamma_{\mu_{1}}\gamma_{\mu_{2}}\slashed{\ku}
  \,u(\kb)\times
  \ubar(\kd)\, \gamma_{6}
  \gamma_{\mu_{1}}\gamma_{\mu_{2}}\slashed{\kt} \,u(\ku)\nonumber \\
  \S7&=&\ubar(\kt)\, \gamma_{6}
  \gamma_{\mu_{1}}\gamma_{\mu_{2}}\gamma_{\mu_{3}}
  \,u(\kb)\times
  \ubar(\kd)\, \gamma_{6}
  \gamma_{\mu_{1}}\gamma_{\mu_{2}}\gamma_{\mu_{3}} \,u(\ku)\nonumber \\
  \S8&=&\ubar(\kt)\, \gamma_{7}
  \gamma_{\mu_{1}}\gamma_{\mu_{2}}\gamma_{\mu_{3}}\slashed{\ku}
  \,u(\kb)\times
  \ubar(\kd)\, \gamma_{6}
  \gamma_{\mu_{1}}\gamma_{\mu_{2}}\gamma_{\mu_{3}} \,u(\ku)\nonumber \\
  \S{9}&=&\ubar(\kt)\, \gamma_{7}
  \gamma_{\mu_{1}}\gamma_{\mu_{2}}\gamma_{\mu_{3}}\gamma_{\mu_{4}}
  \,u(\kb)\times
  \ubar(\kd)\, \gamma_{6}
  \gamma_{\mu_{1}}\gamma_{\mu_{2}}\gamma_{\mu_{3}}\gamma_{\mu_{4}}\slashed{\kt}
  \,u(\ku)\nonumber \\
  \S{10}&=&\ubar(\kt)\, \gamma_{6}
  \gamma_{\mu_{1}}\gamma_{\mu_{2}}\gamma_{\mu_{3}}\gamma_{\mu_{4}}\slashed{\ku}
  \,u(\kb)\times
  \ubar(\kd)\, \gamma_{6}
  \gamma_{\mu_{1}}\gamma_{\mu_{2}}\gamma_{\mu_{3}}\gamma_{\mu_{4}}\slashed{\ku}
  \,u(\ku)\nonumber \\
  \S{11}&=&\ubar(\kt)\, \gamma_{6}
  \gamma_{\mu_{1}}\gamma_{\mu_{2}}\gamma_{\mu_{3}}\gamma_{\mu_{4}}
  \gamma_{\mu_{5}}\,u(\kb)\times
  \ubar(\kd)\,
  \gamma_{6}
  \gamma_{\mu_{1}}\gamma_{\mu_{2}}\gamma_{\mu_{3}}\gamma_{\mu_{4}}
  \gamma_{\mu_{5}}
  \,u(\ku) \, ,
  \label{eq:SpinorStructures}
\end{eqnarray} 
where $\gamma_6=1+\gamma_5$ and $\gamma_7=1-\gamma_5$.
Here, we have not applied any four-dimensional Fierz identities,
which could be used to simplify the structures further
in four dimensions.
We conclude that each of the colour-stripped amplitudes
$B_h$, $B_l$, $A^{(2)}_{1,LC}$ and $A^{(2)}_{1,SC}$
has a decomposition in spinor structures of the form
\begin{equation}
  \label{eq:SpinorDecomposition}
  A^{(2)} = \sum_{i=1}^{11} f_i \S{i} \, .
\end{equation}
Note that the general form of this decomposition
does not depend on the actual values of the scalar
coefficients $C_i$ in \Eq{eq:basictensor}.

We have used two different methods to calculate the scalar coefficients.
The first method, developed by
Tarasov~\cite{Tarasov:1996br,Tarasov:1997kx}, relates these coefficients
to scalar integrals with increased powers of the propagators $\nu_i$
and in higher space-time dimensions. Our implementation of this
method closely follows the approach of \Ref{Anastasiou:1999bn},
where all formulae required for two-loop applications can be found.
Here, we limit ourselves to a very brief summary.
Introducing a Schwinger parameter $x_i$ for each of the propagators
$D_i$, the denominator of \Eq{eq:tensorintegraldef} is written as
\begin{equation}
\frac{1}
{ {D_1}^{\nu_1} \cdots {D_n}^{\nu_n} }
= 
\int {\cal D}x \exp\left(\sum_{i=1}^n x_i D_i \right)\, ,
\end{equation}
where
\begin{equation}
\label{eq:SchwingerDx}
   {\cal D}x =  \prod_{i=1}^n
     \frac{(-1)^{\nu_i} x_i^{\nu_i -1}}{\Gamma(\nu_i)} dx_i\, .
\end{equation}
In terms of the loop momenta, the exponent has the form
\begin{equation}
  \sum_{i=1}^n x_i D_i = a\,\ell_1^2 + b\,\ell_2^2 + 2c\,(\ell_1\cdot\ell_2)
  + 2\,(d\cdot\ell_1) + 2\,(e\cdot\ell_2) + f ,
\end{equation}
with coefficients $a, b, c, d^{\mu}, e^{\mu}, f$ that are linear
in the Schwinger parameters. The loop momentum integrations are
now Gaussian and can be performed straightforwardly. In the case of a
scalar integral, one finds that
\begin{equation}
\label{eq:gaussianQP}
I_n(d,\nu_1,\ldots,\nu_n)[1]
\propto \int {\cal D}x \,
\frac{ \exp\left({\cal Q}/{\cal P}\right) }{ {\cal P}^{d/2} } \, ,
\end{equation}
where ${\cal P}$ and ${\cal Q}$ are polynomials in the Schwinger parameters,
defined by
\begin{align}
{\cal P} & = a b-c^2
\\
{\cal Q} & = -a\,e^2 - b\,d^2 + 2 c\,(d\cdot e) + f {\cal P} \, .
\end{align}
For tensor integrals, similar expressions are obtained,
with
additional polynomials in the Schwinger parameters, divided
by powers of ${\cal P}$, multiplying the integrand in \Eq{eq:gaussianQP}.
From the form of \Eq{eq:SchwingerDx}, one sees that any
additional factor of $x_i$ in the numerator of \Eq{eq:gaussianQP}
can be interpreted as a shift $\nu_i \to \nu_i+1$ in the power of
the corresponding propagator. Similarly, any additional factor
of $1/{\cal P}$ is equivalent to a shift of the
space-time dimension $d \to d+2$. The final result is that
each scalar coefficient $C_i$ in \Eq{eq:basictensor} is expressed as
a sum of scalar integrals $I(d+2m,\nu_1+\delta_1\ldots,\nu_n+\delta_n)[1]$.

In the second method of determining the scalar coefficients, both
sides of \Eq{eq:basictensor} are contracted with tensors $T_i$.
In this way, a system of linear equations is obtained relating
the $C_i$ to integrals with numerators containing scalar
polynomials in the loop momenta. The number of independent
coefficients to be determined can be reduced by exploiting the
symmetry properties of the tensor integrals in \Eq{eq:tensorintegraldef}
with respect to permutations of the Lorentz indices~\cite{Pak:2011xt}.
Both methods lead to identical results in the end, once all scalar
integrals involved have been expressed in terms of a set of independent
master integrals. However, the scalar integrals appearing in the intermediate
steps of the two methods are different.

In addition to these two methods, we have also used a third approach,
where we do not use the scalar coefficients $C_i$ explicitly, but,
instead, directly project out the functions $f_i$ in the spinor
decomposition of \Eq{eq:SpinorDecomposition}. This turns out to be the
most efficient method, especially for the most complicated diagrams,
which are the double boxes.
The projection works by contracting the amplitude with
the conjugate spinor structure $\S{j}^{\ast}$ and taking
the trace of the two spinor chains:
\begin{equation}
  \sum_{i}  \Tr[\S{j}^{\ast}\S{i}] f_i = \Tr[ \S{j}^\ast A^{(2)}].
\end{equation}
We define a matrix of contractions
\begin{equation}
  {\cal M}_{ji} = \Tr[\S{j}^{\ast}\S{i}] \, ,
\end{equation}
whose entries are polynomials in $d$, $\hat{s}$, $\hat{t}$, and $m_t$.
The functions $f_i$ are now obtained from the inverse matrix:
\begin{equation}
  \label{eq:TensorReductionII}
  f_i = \left({\cal M}^{-1}\right)_{ij}\Tr[ \S{j}^\ast A^{(2)}].
\end{equation}
Having proven the general decomposition shown in
\Eq{eq:SpinorDecomposition} with the spinor structures shown in
\Eq{eq:SpinorStructures} we can simplify the calculation of the $f_i$
by setting $\gamma_5$ to zero.  Evidently this has to be done in the
evaluation of ${\cal M}$ and in the evaluation of $\Tr[ S_j^\ast
A^{(2)}]$. This is possible because in a theory with purely vector
couplings we would get precisely the same spinor structures as given
in \Eq{eq:SpinorDecomposition} but with $\gamma_5$ set to zero
everywhere. In other words the scalar functions $f_i$ multiplying the
11 spinor structures do not care whether we have purely vector coupling
or $(V-A)$ coupling as long as we use an anti-commuting $\gamma_5$.
Note that this does not mean that both theories give the same results
for the two-loop amplitude interfered with the Born amplitude. When we
calculate the interference term we need to keep the $\gamma_5$ in the
spin-structures. A similar method has been used recently in
\Ref{Brucherseifer:2014ama} where it has been argued that $\gamma_5$
can be always anti-commuted to the bracketing spinors where one can
absorb $\gamma_5$ into the spinor for a specific helicity.  
(In principle one could also keep $\gamma_5$ in the aforementioned
steps using a concrete prescription to evaluate the trace, i.e. the
method from \Ref{Korner:1991sx}.) 
The trace
operation in $\Tr[\S{j}^{\ast}\S{i}]$ allows to get rid of all spinor
structures and to express the results in terms of two-loop integrals,
where the loop momenta appear only in scalar products with external
momenta or again loop momenta. It is thus possible to cancel some of
them with corresponding factors in the denominator. To proceed, we need
to express these integrals in terms of master integrals. This reduction
will be discussed in the next section.

\section{Reduction to master integrals}
Depending on how the two-loop amplitude is projected onto the eleven
spin structures, the
starting point for the reduction of the tensor integrals to a small
set of scalar master integrals is different. Using the dimension-shifting
method,
one ends up with scalar integrals with
raised powers of the propagators and in higher space-time dimensions.
Inspecting \Fig{fig:twoloop} and \Fig{fig:LC-topos}, we observe that
the most complicated tensor integrals have rank 4, leading to
scalar integrals in $d+8$ dimensions with the powers of the propagators
increased in total by eight units. On the other hand, using the projection
method of \Eq{eq:TensorReductionII} to determine the scalar functions
$f_i$ multiplying the different spinor structures, we obtain scalar
integrals with irreducible scalar products in the numerator. In this case,
the most complicated integrals are double-box integrals with 4 irreducible
scalar products in the numerator. Employing integration-by-parts (IBP)
identities
\cite{Tkachov:1981wb,Chetyrkin:1981qh}, relations between different
integrals can be established. These relations can then be used to
reduce the appearing integrals to the master integrals.  An
algorithmic solution of the reduction procedure has been given with
the Laporta algorithm \cite{Laporta:2001dd}, in which an
overconstrained system of equations is set up, which is then used to
express all integrals in terms of a few integrals. Different
implementations of the Laporta algorithm are publicly available as
computer code \cite{Anastasiou:2004vj, Smirnov:2008iw,%
  Smirnov:2013dia, Studerus:2009ye, vonManteuffel:2012np, Lee:2012cn}.
As the main working horse for the reduction described in this article,
we used the program {\tt Reduze~2}~\cite{vonManteuffel:2012np} written in
C++. In addition, we used a private version of the program {\tt Crusher}
made available to us by P.~Marquard~\cite{crusher}.

\subsection{Reduction of vertex diagrams}
For the reduction of the vertex diagrams we used both methods to
reduce the amplitude to the spinor structures. The highest tensor
integrals are of rank four. As mentioned before, using the
dimension-shifting method
leads
thus to scalar integrals with the powers of the propagators raised by
up to 8 units. In addition, the dimension of the space-time is shifted
to $d+8$. Reducing these integrals using the IBP equations requires
two steps. In the first step, integrals with raised powers of the
propagators defined for arbitrary but fixed space-time dimension are reduced to
master integrals. In a second step, the master integrals in $d+2m$
dimensions ($m=1,\ldots,4$) are transformed back to master integrals
in $d$ dimensions, using the dimension
shift operator $\cal P$
(see \Eq{eq:gaussianQP})
and the
reduction table derived in the first step. Using instead
\Eq{eq:TensorReductionII} to reduce the amplitude to the spin
structures $\S{i}$, we apply the Laporta algorithm to integrals
involving irreducible scalar products. Since all the integrals stay in
$d$ dimensions a dimension shift is not required. For the vertex
corrections, we have applied both methods and found complete
agreement between them.

As a further check on our programs, we have recalculated the two-loop
QCD corrections to the heavy quark vector and axial vector
form factors, applying both the dimension-shifting method and the
projection method. After substituting $\epsilon$-expansions
for the master integrals from \Ref{Gluza:2009yy},
we found complete agreement with the results available in the
literature~\cite{Bernreuther:2004ih,Bernreuther:2004th}.

\subsection{Reduction of the double-box diagrams}
The reduction of the double-box diagrams involves 9 different
topologies, three planar ones and six non-planar topologies.  For each
topology two different diagrams exist. Since the two diagrams
belonging to the same topology are connected, it is sufficient to
determine the reduction tables only once for each of the nine topologies.

The reduction of the double-box topologies is significantly more
complicated than the vertex corrections. The increased
complexity is due to the larger number of propagators and to the
simple fact that the double-box diagrams involve more scales. In the
double-box topologies the $W$-boson mass appears in the two-loop
integrals. In addition, the double-box topologies involve $\hat s$ and
$\hat t$ while the two-loop form factors depend only on $\hat t$.
Applying the dimension-shifting method,
we were not able to generate all the
required reduction tables, using the aforementioned programs.
On the other hand, for the simpler double-box topologies, we were able
to reduce also the highest tensor ranks appearing in the calculation
when applying the projectors of \Eq{eq:TensorReductionII}.
However, even in this case, we were not
able to fully reduce the most complicated double-box topologies. As
mentioned before, the increased complexity of the double-box
topologies as compared to the vertex corrections, is a direct consequence
of the large number of independent variables. In the reduction
rational functions in the five variables $\hat s,\hat t, m_{t},
m_{W}$ and $d$ are generated. The manipulation of these rational
functions, in combination with the related increase of expression size,
leads to much longer run time and also increased memory
consumption. Having said this, the direction to simplify the reduction
is obvious: One needs to reduce the number of independent
variables. Using one variable out of $\hat s,\hat t, m_t, m_W$ to
define the mass scale will reduce the number of independent variables
by one. Expressing in addition the top-quark mass in terms of the
$W$-boson mass (or vice versa), will further reduce the number of
independent variables by one. In \Ref{ATLAS:2014wva} the most precise
measurements from the Tevatron experiments CDF and D0 and the
LHC experiments ATLAS and CMS have been combined.
The {\it world average} quoted in \Ref{ATLAS:2014wva} reads
\begin{equation}
  \label{eq:mtcombined}
  \mt = 173.34 \pm 0.27 \mbox{ (stat) }\pm 0.71 \mbox{ (syst) }
  \mbox{GeV}/c^2.
\end{equation}
Using in addition the world average for $m_W$ as produced by the
particle data group \cite{Beringer:1900zz}
\begin{equation}
  m_{W} = 80.385 \pm 0.015 \;\mbox{GeV}/c^2, 
\end{equation}
we can set to very good approximation
\begin{equation}
  m_t^2 \approx {14\over 3} m_W^2,\label{eq:mt-mw-ratio}
\end{equation}
which is equivalent to 
\begin{equation}
  m_t \approx 173.65\;\mbox{GeV}/c^2. 
\end{equation}
This value for the top-quark mass is compatible with the aforementioned
world average and deviates less than 2 per mille from the central
value quoted in \Eq{eq:mtcombined}. Since single top-quark production
depends very weakly on the top-quark mass around the nominal value,
the uncertainty introduced by the aforementioned approximation is
completely negligible. (Using this approximation in leading
order would lead to effects at the per mille level. For more details
we refer to \Ref{Kant:2014oha} where the mass dependence of single
top-quark production has been studied in detail.) Choosing \Eq{eq:mt-mw-ratio}
to reduce the number of independent variables,
leads indeed to an enormous simplification of the reduction procedure.
In addition, we also fine tuned, for the most complicated topologies,
the seed generation in the Laporta algorithm. Using these two
techniques, we were able to reduce all the double-box integrals to
master integrals. In cases where the reduction was feasible for
arbitrary values of $m_W$ and $m_t$, the two approaches agreed, after
specializing the general results to the specific case
$m_t^2= {14\over 3} m_W^2$. In addition, we observed a dramatic reduction
in the size of the final expressions, when setting $m_t^2= {14\over 3} m_W^2$.

\section{Results}
In this section, we present analytic results for $A_{1,LC}^{(2)}$,
$B_h$ and $B_l$. The vertex contributions to $A_{1,SC}^{(2)}$ are
given in appendix \ref{sec:A1SCvertex}.
The master integrals entering the corrections to the $W$-$t$-$b$ vertex are
known in the literature from studies of the form factors describing the
decay $b \to u + W^*$, with $m_u=0$, $m_b\neq 0$
\cite{Bell:2006tz,Bell:2007tv,Bell:2009nk,Bonciani:2008wf,Beneke:2008ei,%
Asatrian:2008uk,Huber:2009se}.
For master integrals entering the corrections to the light quark vertex,
see also \Ref{Bonciani:2003hc}.
In the results shown below, the master integrals are kept as
symbols, and also the full $d$ dependence is kept.
In choosing the basis for the master integrals, we follow
\Refs{Bell:2006tz,Bell:2007tv,Bell:2009nk}.
The definitions of the master integrals are given in
appendix \ref{sec:masterintegrals}.
For the presentation of the results, it is convenient to
introduce rescaled invariants
\begin{equation}
  t = {\hat t\over \mt^2},\quad   
  s = {\hat s\over \mt^2},
\end{equation}
and rescaled spinor structures
\begin{equation}
  \Shat1 = {\S1 \over \mt},\quad   
  \Shat3 = \S3.
\end{equation}
Note, that the vertex corrections only depend on the $W$-boson mass
through the $W$-boson propagator. All vertex contributions thus have a
universal factor 
\begin{displaymath}
  {1\over \hat t -\mw^2}.
\end{displaymath}
Furthermore, the form factors only depend on $\hat t$. The entire $\hat s$
dependence is thus hidden in the spinor structures $\Shat{1}$ and
$\Shat{3}$.
For the contribution of the light quark loops we obtain: 
\def\MIOneZeroOnep{\text{MI101p}}
\def\MITwoZeroOne{\text{MI201}}
\def\MITwoZeroOnep{\text{MI201p}}
\def\MIThreeZeroOne{\text{MI301}}
\def\MIThreeZeroOnep{\text{MI301p}}
\def\MIThreeZeroOnepuOne{\text{MI301pu1}}
\def\MIThreeZeroTwop{\text{MI302p}}
\def\MIThreeZeroThreep{\text{MI303p}}
\def\MIFourZeroOne{\text{MI401}}
\def\MIFourZeroTwop{\text{MI402p}}
\def\MIFourZeroSixt{\text{MI406t}}
\def\MIFourZeroSevenp{\text{MI407p}}
\def\MIFourZeroSevent{\text{MI407t}}
\def\MIFourZeroEightp{\text{MI408p}}
\def\MIFiveZeroOnep{\text{MI501p}}
\def\MIFiveZeroTwop{\text{MI502p}}
\begin{align}
2 (\hat{t} - m_{W}^2) \times & B_l=  \hspace{-4mm}\notag \\ 
  &\hspace{-16mm}\frac{\MIThreeZeroOnepuOne}{16 (d - 4) (d - 1) (2 d - 7) (t - 1)} (d - 2)  \notag \\
 \times
   \Bigl[
    \Shat{1}&
  \Bigl( 
  16 (d^3 - 11 d^2 + 39 d - 44)
   \Bigr) 
-  \Shat{3}
   \Bigl( 
   (3 d - 8)  (d (t - 3) - 6 t + 10)
     \Bigr)
 \Bigr] \notag \\
  &\hspace{-16mm}-\frac{\MIFourZeroOne}{ 2 (d - 4) (d - 1) (3 d - 8)} \notag \\
\times
 \Bigl[
  \Shat{3}&
\Bigl( 
(d - 2) (3 d^3 - 31 d^2 + 110 d - 128)
\Bigr)
 \Bigr] \notag \\
  &\hspace{-16mm}-\frac{\MIFourZeroTwop}{2 (d - 4) (d - 1) (3 d - 8) (t - 1)}  \notag \\
\times    
\Bigl[
    \Shat{1}&
  \Bigl( 
  8 (d - 4) (d - 2) (d^2 - 7 d + 11)
  \Bigr) \notag \\
 + \Shat{3}&
  \Bigl( 
  (d - 2) (3 d^3 - 31 d^2 + 110 d - 128) t - (d - 4) (d - 2) (3 d^2 - 
    22 d + 36)
  \Bigr)
\Bigr] \, .
\end{align}

The contribution due to closed top-quark loops reads:
\begin{align}
\hspace{-5mm}2 (\hat{t} - m_{W}^2) \times & B_h=\notag \\ 
  &\hspace{-16mm}\frac{(\MIOneZeroOnep)^2}{16 (d - 7)^2 (d - 6) (d - 5)^2 (d - 4) (d - 3) (d - 1) (3 d - 8) (t - 
   1)^3 t} \notag \\
    \times \Bigl[
 4 \Shat{1} &(d - 7)^2 (d - 2)
 \Bigl(
 - (d - 2) (11 d^6 - 270 d^5 + 2670 d^4 \notag \\ 
 & -
 13592 d^3 +  37443 d^2 - 52686 d + 29424) t^3 \notag \\ 
 & +
2 (17 d^7 - 466 d^6 + 5294 d^5 - 32312 d^4 +  
 114473 d^3 - 235626 d^2 + 261228 d - 120408) t^2 \notag \\ 
 & +
(17 d^7 - 424 d^6 + 4414 d^5 - 24884 d^4 + 
 82377 d^3 - 161352 d^2 + 174804 d - 81552) t
 \Bigr)\notag \\        
       + 
 \Shat{3} &(d - 2)
 \Bigl(
 -2 (d - 7) (d - 6)^2 (d - 4) (d - 3) (d - 2) (d - 1) (3 d -  8) t^5 \notag \\ 
 & +
 (7 d^8 - 256 d^7 + 4017 d^6 - 35257 d^5 +  189540 d^4 \notag \\ 
 & -
 641323 d^3 + 1342244 d^2 - 1604508 d +   847776) t^4 \notag \\ 
 & +
 (24 d^9 - 999 d^8 + 18280 d^7 -    192775 d^6 + 1288447 d^5 - 5644750 d^4 \notag \\ 
 & +
 16162933 d^3 - 29085772 d^2 + 29772324 d - 13172352) t^3 \notag \\ 
 & +
 (209 d^8 -  7096 d^7 + 103443 d^6 - 843419 d^5 + 4192720 d^4 \notag \\ 
 & -
 12956457 d^3 +  24167772 d^2 - 24676020 d + 10419168) t^2 \notag \\ 
 & -
 (d - 3)  (24 d^8 - 781 d^7 + 10663 d^6 - 79156 d^5 + 347337 d^4 \notag \\ 
 & -
 926691 d^3 + 1529312 d^2 - 1610140 d + 963072) t \notag \\ 
 & +
 32 (d - 7)^2 (d - 6) (d - 5) (d - 4) (d - 3)^2  (d - 1)
 \Bigr)
    \Bigr] \notag \\   
 &\hspace{-16mm}{}+\frac{\MIOneZeroOnep\times\MITwoZeroOne}{(d - 7) (d - 5) (d - 4) (d - 1) (3 d - 8) t}\notag \\ \times 
 \Bigl[
 \Shat{3}&
   \Bigl( 
    (d^2 - 5 d + 6) (3 d^3 - 29 d^2 + 74 d - 48) t \notag \\ 
 & -
    2 (d^2 - 5 d + 6) (3 d^4 - 51 d^3 + 334 d^2 - 946 d + 936)
   \Bigr)
    \Bigr] \notag \\  
 &\hspace{-16mm}{} +\frac{\MIOneZeroOnep\times\MITwoZeroOnep}{(d-7) (d-5) (d-4) (d-1) (3 d-8) (t-1)^3}\notag \\ \times 
 \Bigl[
  \Shat{3}&
   \Bigl(
   (d - 6) (d - 3) (d - 2) (d - 1) (3 d - 8) t^3 \notag \\ 
 & -
   2 (d - 3) (d - 2) (3 d^4 - 48 d^3 + 305 d^2 - 872 d +   888) t^2 \notag \\ 
 & +
   (d - 5) (d - 3) (d - 2) (6 d^3 - 81 d^2 + 338 d - 416) t
   \Bigr) \notag \\ 
  +   \Shat{1}&  
   \Bigl(
   -8 (d - 7) (d - 4) (d - 3) (d - 2) (d^2 - 12 d + 21)  t
   \Bigr)
    \Bigr] \notag \\   
 &\hspace{-16mm}{} +\frac{\MIThreeZeroThreep}{8 (d-7) (d-6) (d-4) (d-1) (t-1)^3} \notag \\ \times 
\Bigl[
  \Shat{3}&
  \Bigl(
  -(3 d - 8) (2 d^3 - 27 d^2 + 93 d - 58) t^3 \notag \\ 
 & +
  (4 d^4 - 49 d^3 +  133 d^2 + 62 d - 320) t^2 \notag \\ 
 & -
  (3 d - 8) (2 d^3 - 21 d^2 + 55 d +  2) t \notag \\ 
 & +
  (d - 3) (8 d^3 - 103 d^2 + 386 d - 256)
  \Bigr)\notag \\ 
  +  4 \Shat{1}&  (d - 7) 
   \Bigl(
(d^4 - 14 d^3 + 59 d^2 - 82 d + 16) t^2 \notag \\ 
 & -
 2 (d - 2) (3 d^3 - 30 d^2 + 73 d - 26) t \notag \\ 
 & -
   (3 d^4 - 30 d^3 + 89 d^2 - 98 d + 56)
   \Bigr)
 \Bigr] \notag \\ 
     &\hspace{-16mm}{} +\frac{\MIFourZeroSixt}{4 (d-7) (d-5) (d-1) (3 d-8) t}\notag \\ \times 
 \Bigl[
  \Shat{3}&
  \Bigl(   
  (d - 6) (d - 1) (3 d - 10) (3 d - 8) t^2 \notag \\ 
 & -
  2 (d - 4) (3 d^4 - 55 d^3 + 334 d^2 - 674 d + 300) t + 
 32 (d - 7) (d - 5) (d - 3) (d - 1)
    \Bigr)
 \Bigr] \notag \\ 
  &\hspace{-16mm}{} +\frac{\MIFourZeroSevenp}{4 (d - 7) (d - 5) (d - 1) (3 d - 8)
 (t-1)^3}\notag \\ \times 
 \Bigl[
  \Shat{1}&
  \Bigl(   
  8 (d - 7) (d - 4) (d^3 - 2 d^2 + d - 10) t^2 \notag \\ 
 & +
 16 (d - 7) (3 d^4 - 34 d^3 + 119 d^2 - 138 d + 20) t \notag \\ 
 & +
 8 (d - 7) (d - 4) (d^3 - 2 d^2 + d - 10)
  \Bigr)\notag \\ 
  +  \Shat{3}&  
  \Bigl( 
  (d - 6) (d - 1) (3 d - 10) (3 d - 8) t^4 \notag \\ 
 & -
 2 (d - 5) (3 d^4 - 34 d^3 + 150 d^2 - 236 d + 48) t^3 \notag \\ 
 & +
 2 (d - 4) (9 d^4 - 116 d^3 + 491 d^2 - 630 d - 40) t^2 \notag \\ 
 & -
 2 (9 d^5 - 155 d^4 + 972 d^3 - 2582 d^2 + 2324 d + 320) t \notag \\ 
 & +
 (d -   5) (2 d^2 - 17 d + 32) (3 d^2 - 16 d + 4)
  \Bigr)
 \Bigr] \notag \\   
  &\hspace{-16mm}{} +\frac{\MIFourZeroSevent}{2 (d - 7) (d - 5) (d - 1) (3 d - 8)
}\notag \\ \times 
 \Bigl[
  \Shat{3}&
  \Bigl(    
  (d - 6) (d - 1) (3 d - 8) t^2 + 2 (5 d^3 - 45 d^2 + 64 d + 68) t - 
 8 (d^3 - 13 d^2 + 84 d - 164)
 \Bigr)
 \Bigr] \notag \\ 
  &\hspace{-16mm}{} +\frac{\MIFourZeroEightp}{2 (d-7) (d-5) (d-1) (3 d-8) (t-1)}\notag \\ \times 
 \Bigl[
  \Shat{1}&
  \Bigl( 
  8 (d - 7) (d^3 - 8 d^2 + 23 d - 26) + 
 8 (d - 7) t (d^3 - 8 d^2 + 23 d - 26)
  \Bigr)\notag \\ 
  +  \Shat{3}&  
  \Bigl( 
  (d - 6) (d - 1) (3 d - 8) t^3 + 
  (-5 d^3 + 55 d^2 - 242 d +376) t^2 \notag \\ 
 & -
  (3 d - 8) (d^2 - 3 d - 2) t + (d - 5) (d - 4) (5 d -   22)
    \Bigr) 
 \Bigr] \, .
 \end{align}
The leading colour contribution is given by:
\begin{align}
(\hat{t} - m_{W}^2) &\times A^{(2)}_{1,LC}=\hspace{-16mm}\notag \\ 
 &\hspace{-16mm}{} \frac{(\MIOneZeroOnep)^2}{32 (d - 5) (d - 4)^2 (d - 3)^2 (t - 1)^3 t}  (d - 2)
 \notag \\ \times
  \Bigl[2 \Shat{1}& (d -4) 
   \Bigl(
   -(d - 2) (2 d^5 - 35 d^4 + 244 d^3 - 779 d^2 + 980 d - 144) t^3  \notag \\        
 &
   + (8 d^6 - 157 d^5 + 1140 d^4 - 3993 d^3 + 6782 d^2 - 3984 d - 1184) t^2  \notag \\        
 &
   - (6 d^6 - 117 d^5 + 950 d^4 - 3821 d^3 + 7394 d^2 - 5300 d - 380) t \notag \\        
 &
   + (d - 3)^2 (d^3 - 14 d^2 + 60 d - 84)
   \Bigr) \notag \\ 
  +   \Shat{3}&   
   \Bigl(
   -(d - 2) (5 d^4 - 53 d^3 + 182 d^2 - 186 d -  48) t^4  \notag \\        
 &
   + (4 d^7 - 89 d^6 + 825 d^5 - 4036 d^4 + 10626 d^3 - 12526 d^2 - 588 d + 10112) t^3  \notag \\        
 &
   - (8 d^7 - 185 d^6 + 1747 d^5 - 8434 d^4 + 20774 d^3 - 19174 d^2 - 13504 d +  28112) t^2 \notag \\        
 &
   + (4 d^7 - 95 d^6 + 903 d^5 - 4228 d^4 +  9254 d^3 - 3874 d^2 - 17524 d + 21120) t \notag \\        
 &
   - (d - 4) (d - 3)^2 (d^3 - 14 d^2 + 60 d - 84)
   \Bigr)
 \Bigr] 
 \notag \\ 
   &\hspace{-16mm}-{} \frac{ \MIOneZeroOnep\times\MITwoZeroOne}{4 (d - 4)^2 (d - 3) (t - 1)} \times(d^3 - 9 d^2 + 30 d - 32) \times
   \big[
   \Shat{1} (d^2 - 9 d + 20)  + \Shat{3} (d + t - 3)
   \bigr]  \notag \\ 
  &\hspace{-16mm}+{} \frac{\MIOneZeroOnep\times\MITwoZeroOnep}{8 (d - 5) (d - 4)^2 (d - 3) (t - 1)^3 (t + 1)} \notag \\ \times 
 \Bigl[
  2  \Shat{1} & (d - 4) 
  \Bigl(
  (d - 2) (6 d^4 + 6 d^3 - 599 d^2 + 2965 d -  4194) t^3 \notag \\ 
 & -
  (2 d^6 - 132 d^5 + 1624 d^4 - 8801 d^3 +  25103 d^2 - 37864 d + 24316) t^2 \notag \\ 
 & +
  (50 d^5 - 618 d^4 + 2823 d^3 - 5771 d^2 + 4952 d - 1140) t \notag \\ 
 & +
  (2 d^6 - 44 d^5 + 424 d^4 - 2301 d^3 + 7367 d^2 -  12988 d + 9676)
  \Bigr) \notag \\ 
  + \Shat{3} &   
 \Bigl(
 -(d - 2) 
 (d^5 - 33 d^4 + 361 d^3 - 1811 d^2 + 4330 d - 4040) t^4 \notag \\ 
 & +
 (-2 d^7 + 58 d^6 - 723 d^5 + 5167 d^4 - 23052 d^3 +  63656 d^2 - 99048 d + 65856) t^3 \notag \\ 
 & +
 (2 d^7 - 68 d^6 + 1001 d^5 -  8085 d^4 + 38330 d^3 - 106292 d^2 + 159488 d -  99936) t^2 \notag \\ 
 & +
 (2 d^7 - 42 d^6 + 307 d^5 - 815 d^4 - 548 d^3 +  6520 d^2 - 9560 d + 2496) t \notag \\ 
 & -
 (d - 4) (2 d^6 - 45 d^5 + 440 d^4 -  2400 d^3 + 7663 d^2 - 13416 d + 9916)
 \Bigr)
  \Bigr] 
 \notag \\ 
  &\hspace{-16mm}+{} \frac{ (\MITwoZeroOne)^2}{4 (d - 4)^2}\times
  \big[
  \Shat{3} (d^2 - 7 d + 16)^2 
   \bigr]  \notag \\ 
  &\hspace{-16mm}+{} \frac{ \MITwoZeroOne\times\MITwoZeroOnep }{4 (d - 4)^2 (t - 1)} \times  (d^2 - 7 d + 16) \notag \\ \times 
 \Bigl[
   2 \Shat{1} & (d - 5) (d - 4)  + 
     \Shat{3} \bigl((d^2 - 7 d + 16) t - (d - 5) (d - 4)\bigr)
  \Bigr]  \notag  \\  
   &\hspace{-16mm}+{} \frac{(\MITwoZeroOnep)^2 }{4 (d - 4)^2 (d - 2) (t - 1)^2}\notag \\ \times 
 \Bigl[
  2 \Shat{1} & (d - 5) (d - 4)  
  \Bigl((d - 2) (d^2 - 7 d + 20) t - (d - 4)^2 (d - 3)\Bigr) \notag \\ 
  + \Shat{3} &   
 \Bigl(
 (d - 5) (d - 3) (d - 4)^3 - 2 (d - 5) (d^3 - 9 d^2 + 31 d - 36) t (d - 4) \notag \\ 
 & +
 (d - 2) (d^2 -  7 d + 16)^2 t^2
 \Bigr)
  \Bigr]  \notag  \\ 
    &\hspace{-16mm}-{} \frac{\MIThreeZeroOne }{2 (d - 4)^3 (d - 3) t}\times
  \Shat{3}(3 d - 8) (d^5 - 18 d^4 + 138 d^3 - 552 d^2 + 1144 d - 980) \notag  \\ 
    &\hspace{-16mm}+{} \frac{\MIThreeZeroOnep }{48 (d - 4)^2 (d - 3) (d - 2) (3 d - 14) (3 d - 10) (t - 1)^4 t (t + 1)}
    (3 d - 8) \notag \\ \times 
 \Bigl[
  2 \Shat{1} &(d -  4)  
  \Bigl(
  -3 (d - 2) (39 d^5 - 168 d^4 - 2953 d^3 + 26640 d^2 -  78364 d + 79264) t^5 \notag \\ 
 & -
  (1413 d^6 - 27981 d^5 + 234094 d^4 - 1060948 d^3 + 2745592 d^2 - 3833952 d + 2244096) t^4 \notag \\ 
 & +
  2 (1206 d^6 - 21423 d^5 + 162301 d^4 - 676348 d^3 + 1640716 d^2 - 2192784 d + 1253568) t^3 \notag \\ 
 & +
  4 (459 d^6 - 7374 d^5 + 47544 d^4 - 157328 d^3 + 283267 d^2 -   268548 d + 110124) t^2 \notag \\ 
 & -
  (135 d^6 - 3516 d^5 + 39101 d^4 -  229022 d^3 + 734948 d^2 - 1221192 d + 821280) t \notag \\ 
 & +
  (10 - 3 d)^2 (d - 3) (d^3 - 14 d^2 + 60 d - 84)
     \Bigr)  \notag  \\    
  +\Shat{3} & 
  \Bigl(
  3 (d - 2) (3 d - 10) (3 d^5 - 82 d^4 + 825 d^3 - 3946 d^2 +   9128 d - 8272) t^6 \notag \\ 
 & -
  (279 d^7 - 7464 d^6 + 87645 d^5 -  581978 d^4 + 2342908 d^3 \notag \\ 
 & -
 5677752 d^2 + 7621632 d -  4349568) t^5 \notag \\ 
 & +
  (837 d^7 - 22503 d^6 + 264520 d^5 -  1748672 d^4 + 6969480 d^3 \notag \\ 
 & -
 16643600 d^2 + 21946528 d -  12284544) t^4 \notag \\ 
 & -
  2 (387 d^7 - 11778 d^6 + 147097 d^5 - 987170 d^4 + 3867090 d^3 \notag \\ 
 & -
 8873800 d^2 + 11068952 d - 5798304) t^3 \notag \\ 
 & +
  (9 d^7 - 2208 d^6 +     36717 d^5 - 247402 d^4 + 816056 d^3 \notag \\ 
 & -
 1258312 d^2 + 540464 d +  376704) t^2 \notag \\ 
 & +
  (189 d^7 - 5676 d^6 + 71951 d^5 - 498558 d^4 +   2039584 d^3 \notag \\ 
 & -
 4928840 d^2 + 6519088 d - 3642816) t \notag \\ 
 & -
  (10 -  3 d)^2 (d - 4) (d - 3) (d^3 - 14 d^2 + 60 d - 84)
  \Bigr)
   \Bigr]  \notag  \\ 
  &\hspace{-16mm}+{} \frac{\MIThreeZeroOnepuOne }{32 (d - 4)^2 (d - 3) (d - 2) (d - 1) (2 d - 7) (t - 1)^3 t}\notag \\ \times 
 \Bigl[
2 \Shat{1} & (d - 4) 
\Bigl(
-(d - 2) (d^6 + 440 d^5 - 6429 d^4 + 35932 d^3 -    95824 d^2 + 119592 d - 53472) t^3 \notag \\ 
 & -
(529 d^7 - 9252 d^6 +   70159 d^5 - 299288 d^4 + 772684 d^3 - 1195216 d^2 +  1007616 d 
\notag \\ 
 & 
- 346752) t^2 \notag \\ 
 & -
(d - 3) (199 d^6 - 1941 d^5 +    4754 d^4 + 9760 d^3 - 63060 d^2 + 97640 d - 47232) t \notag \\ 
 & +
(d -  3) (d - 1) (3 d - 10) (3 d - 8) (d^3 - 14 d^2 + 60 d -  84)
\Bigr)  \notag \\        
+\Shat{3} & (3 d - 8) 
\Bigl(
-(d - 2) (20 d^5 - 311 d^4 + 1903 d^3 - 5618 d^2 + 7740 d - 3704) t^4 \notag \\ 
 & -
(31 d^7 - 704 d^6 + 7086 d^5 -    40145 d^4 + 135102 d^3 - 263900 d^2 + 269960 d -   107328) t^3 \notag \\ 
 & +
(59 d^7 - 1512 d^6 + 16118 d^5 - 92351 d^4 +   305548 d^3 - 578716 d^2 + 572296 d - 221328) t^2 \notag \\ 
 & -
(d -    4) (25 d^6 - 652 d^5 + 5982 d^4 - 26367 d^3 + 60042 d^2 -   66876 d + 27832) t \notag \\ 
 & -
(d - 4) (d - 3) (d - 1) (3 d - 10) (d^3 -  14 d^2 + 60 d - 84)
\Bigr)
  \Bigr]  \notag  \\ 
  &\hspace{-16mm}+{} \frac{\MIThreeZeroTwop }{48 (d - 4)^2 (d - 3)^2 (d - 2) (3 d - 14) (3 d - 10) (t - 1)^4 t (t +  1)}\notag \\ \times 
 \Bigl[
  2 \Shat{1} &(d - 4)  
  \Bigl(-3 (d - 4) (d - 2) (39 d^5 - 168 d^4 - 2953 d^3 +   26640 d^2 - 78364 d + 79264) t^6 \notag \\ 
 & -
  (774 d^7 - 26721 d^6 +   356317 d^5 - 2478482 d^4 + 9901244 d^3 \notag \\ 
 & -
 22954024 d^2 +    28764096 d - 15080448) t^5 \notag \\ 
 & +
  (10827 d^7 - 223305 d^6 +     1981624 d^5 - 9871708 d^4 + 30022960 d^3 \notag \\ 
 & -
 56104608 d^2 +   59923968 d - 28266240) t^4 \notag \\ 
 & +
  2 (4770 d^7 - 91863 d^6 + 740791 d^5 - 3234600 d^4 +  8241086 d^3 \notag \\ 
 & -
 12240312 d^2 + 9841128 d -   3343104) t^3 \notag \\ 
 & +
  (1377 d^7 - 18996 d^6 + 62875 d^5 +    328634 d^4 - 3249632 d^3 \notag \\ 
 & +
 10583832 d^2 - 16053648 d +   9585216) t^2 \notag \\ 
 & -
  (126 d^7 - 3213 d^6 + 36817 d^5 - 240990 d^4 +    954816 d^3 - 2255336 d^2 + 2907792 d 
\notag \\ 
 & -
 1564800) t \notag \\ 
 & +
  (10 -   3 d)^2 (d - 4) (d - 3) (d^3 - 14 d^2 + 60 d - 84)
    \Bigr) \notag \\ 
  + \Shat{3} &  
 \Bigl(3 (d - 4) (d - 2) 
 (3 d - 10) (3 d^5 - 82 d^4 + 825 d^3 -  3946 d^2 + 9128 d - 8272) t^7 \notag \\ 
 & -
 2 (99 d^8 - 3687 d^7 + 56907 d^6 - 485332 d^5 + 2527836 d^4 -   8280296 d^3 \notag \\ 
 & +
 16711104 d^2 -19033056 d + 9376896) t^6 \notag \\ 
 & +
 (1854 d^8 - 54405 d^7 + 718453 d^6 -   5542690 d^5 + 27121572 d^4 - 85585512 d^3 \notag \\ 
 & +
 169096992 d^2 -  190429312 d + 93318144) t^5 \notag \\ 
 & +
 (-2205 d^8 + 75537 d^7 -   1089466 d^6 + 8726684 d^5 - 42714580 d^4 + 131331008 d^3 \notag \\ 
 & -
 248352496 d^2 + 264603584 d - 121790976) t^4 \notag \\ 
 & +
 (-333 d^8 +   3348 d^7 + 13303 d^6 - 329746 d^5 + 1816836 d^4 -  4238712 d^3 \notag \\ 
 & +
 2610176 d^2 + 5461120 d - 7300608) t^3 \notag \\ 
 & +
 2 (342 d^8 - 12798 d^7 + 192703 d^6 - 1576044 d^5 + 7786640 d^4 -   24034128 d^3 \notag \\ 
 & +
 45561040 d^2 - 48743168 d +  22627584) t^2 \notag \\ 
 & +
 (d - 4) (180 d^7 - 4833 d^6 + 55603 d^5 -  354122 d^4 + 1343364 d^3 \notag \\ 
 & -
 3023192 d^2 + 3720832 d -  1922496) t \notag \\ 
 & -
 (10 - 3 d)^2 (d - 4)^2 (d - 3) (d^3 - 14 d^2 +   60 d - 84)
\Bigr)
  \Bigr] \notag  \\
  &\hspace{-16mm}+{} \frac{\MIFourZeroOne }{4 (d - 4)^2 (d - 1) (3 d - 8)}\notag \\ \times 
 \Bigl[
 \Shat{3}& (-3 d^6 + 82 d^5 - 819 d^4 + 4030 d^3 - 10344 d^2 + 12824 d - 5632)
  \Bigr] \notag  \\
  &\hspace{-16mm}-{} \frac{\MIFourZeroTwop }{(d - 2) (d - 1) (3 d - 8) 8 ( d -4 )^2 (t - 1)^3 (t + 1)} \notag \\ \times 
 \Bigl[
 2 \Shat{1} & (d -4 )  
\Bigl(
 (d - 2) (17 d^5 - 353 d^4 + 2572 d^3 - 8428 d^2 + 12232 d -  5920) t^3 \notag \\ 
 & +
 (-143 d^6 + 2019 d^5 - 12632 d^4 + 43772 d^3 -  85488 d^2 + 85936 d - 33344) t^2 \notag \\ 
 & +
 (-17 d^6 - 141 d^5 +  3278 d^4 - 18348 d^3 + 46172 d^2 - 54120 d + 23296) t \notag \\ 
 & +
 (-d^6 + 45 d^5 - 524 d^4 + 2740 d^3 - 7268 d^2 +  9368 d - 4480)
\Bigr)  \notag \\ 
  + \Shat{3} &  (t - 1)
 \Bigl(
 2 (d - 2) (3 d^6 - 82 d^5 + 819 d^4 - 4030 d^3 + 10344 d^2 -   12824 d + 5632) t^3 \notag \\ 
 & +
 (-33 d^7 + 779 d^6 - 7936 d^5 + 44060 d^4 -  141224 d^3 + 257488 d^2 - 243552 d + 90112) t^2 \notag \\ 
 & + 
 2 (d - 4) (15 d^6 - 356 d^5 + 3033 d^4 - 12592 d^3 + 27374 d^2 -   29524 d + 12096) t \notag \\ 
 & -
 (d - 4) (3 d^6 - 73 d^5 + 684 d^4 -  3212 d^3 + 8004 d^2 - 9912 d + 4608)
\Bigr) 
   \Bigr]  \notag  \\
  &\hspace{-16mm}+{} \frac{\MIFiveZeroOnep}{8 (d - 3) (d - 2) (t - 1)^3 t (t + 1)}\notag \\ \times 
 \Bigl[
  \Shat{1} &
  \Bigl(
 2 (d - 8) (d - 4) (d - 2)^2 t^5 \notag \\ 
 & -
 2 (d - 2) (29 d^3 + 72 d^2 - 1600 d + 3552) t^4 \notag \\ 
 & -
 4 (183 d^4 - 1880 d^3 + 7594 d^2 - 14890 d + 11984) t^3 \notag \\ 
 & -
 4 (97 d^4 - 766 d^3 + 1686 d^2 - 94 d - 2064) t^2 \notag \\ 
 & +
 2 (13 d^4 - 288 d^3 + 1968 d^2 - 5508 d + 5552) t \notag \\ 
 & -
 2 (d - 4) (d^3 - 14 d^2 + 60 d - 84)
  \Bigr) \notag \\ 
  + \Shat{3} &  
 \Bigl(
 -(d - 2) (3 d^3 + 8 d^2 - 148 d + 336) t^5 \notag \\ 
 & +
 (-47 d^4 +    650 d^3 - 4092 d^2 + 12436 d - 13888) t^4 \notag \\ 
 & +
 2 (43 d^4 - 780 d^3 + 5076 d^2 - 14092 d + 13984) t^3 \notag \\ 
 & -
 2 (9 d^4 - 292 d^3 + 2076 d^2 - 5400 d + 4616) t^2 \notag \\ 
 & +
 (-19 d^4 +   346 d^3 - 2188 d^2 + 5904 d - 5856) t \notag \\ 
 & +
 (d - 4) (d^3 -  14 d^2 + 60 d - 84)
\Bigr)
  \Bigr]  \notag  \\
  &\hspace{-16mm}+{} \frac{\MIFiveZeroTwop}{2 (d - 4)^2 (d - 3) (d - 2) t (t^2 - 1)} (2 d - 9)\notag \\ \times 
 \Bigl[
  2 \Shat{1} & (d - 4) 
\Bigl(
  (d - 2) (d^2 + 22 d - 96) t^3 +       
 3
   (11 d^3 - 74 d^2 + 200 d - 208) t^2 \notag \\ 
 & +
  (15 d^3 - 52 d^2 - 40 d + 180) t - 
   (d^3 - 14 d^2 + 60 d - 84)
 \Bigr) \notag \\ 
  + \Shat{3}&  
 \Bigl(
 2 (d - 2)  (2 d^2 - 15 d + 30) t^4 + 
  (5 d^4 -    74 d^3 + 470 d^2 - 1408 d + 1544) t^3 \notag \\ 
 & -
  (9 d^4 -  166 d^3 + 1094 d^2 - 3068 d + 3064) t^2 +
 (d - 2)  (3 d^3 - 72 d^2 + 402 d - 652) t \notag \\ 
 & +
 (d - 4)  (d^3 - 14 d^2 + 60 d - 84)
 \Bigr)
  \Bigr] \, .
\end{align}

The sub-leading colour contribution of the vertex corrections to
$A_{1,SC}^{(2)}$ is given in appendix \ref{sec:A1SCvertex}.  We
have also obtained analytic expressions for the double-box diagrams.
However, the expressions of several megabytes are too long to be presented
here. Since in the end, the main interest is in the numerical
evaluation of the corrections and not so much in the analytic
structure, we have generated computer code written in C to evaluate the
contributions from double-box topologies.
In the functions $f_i$ appearing in the decomposition of the amplitude
according to its spin structure as shown in \Eq{eq:SpinorDecomposition},
we have expanded the coefficient of each master integral
in $\epsilon=(4-d)/2$:
\begin{equation}
  \label{eq:NumImplementation}
  f_i = \sum_{r}\sum_{s=-5}^4 \epsilon^s\, f_{i,r,s}\,\mbox{MI}_r \, .
\end{equation}
The index $r$ labels the master integrals MI. While some of them are
known in the literature, the most complicated double-box integrals are
not yet known and need to be calculated in the future. Assuming that
only ${1\over \epsilon^4}$ poles occur in the master integrals
$\mbox{MI}_r$, it is sufficient to keep terms up to fourth order in
$\epsilon$ in the coefficients of the master integrals. Note that the
reduction procedure itself can lead to spurious poles in $\epsilon$.
These poles require to calculate also positive powers in $\epsilon$
for the master integrals. Alternatively, one may consider to change the
basis of master integrals, to avoid the spurious poles to some extent.
As mentioned in the previous section, we have calculated the double-box
contribution for a fixed ratio $\mw^2/\mt^2$. Using again the rescaled
invariants $s$ and $t$ the coefficients are rational functions in $s$
and $t$ only. As proof of concept we have calculated all coefficients
$f_{i,r,s}$ introduced above. They are encoded in the aforementioned C
library. We have checked that
the numerical results obtained using the C library agree with the original
mathematica code used to produce the expressions. In this comparison
it turns out to be crucial to use extended (quadruple) floating point precision.
We trace this back to the observation, that integer constants occurring
in the numerical evaluation vary over many orders of magnitude.
We note that it is
straightforward to change the integral basis used in the numerical
evaluation. This can be done by rerunning the code generation. The code
for the coefficients as introduced in \Eq{eq:NumImplementation} can be
obtained on demand. Using extended floating point precision in the
evaluation of the coefficients $f_{i,r,s}$ leads to an increased
runtime. We do not consider this as a major problem: First of all, for
the integration of the virtual corrections, typically a small number of phase
space points is usually sufficient. In the practical application, one
may calculate the two-loop contribution as a two-dimensional grid in
$s$ and $t$, which is interpolated during the calculation. The
calculation of the grid can be parallelized. Furthermore, most likely
the computational effort of the complete NNLO calculation will
be dominated by the evaluation of the real corrections, in
particular the double unresolved contributions.

\section{Conclusion}
In this article we consider two-loop QCD corrections for single
top-quark production in the $t$-channel. We have decomposed the
two-loop amplitude according to its colour and spin structure. Using an
anti-commuting $\gamma_5$, eleven different spin structures occur in
the most complicated contributions. To reduce the two-loop tensor
integrals to master integrals, we used the publicly available program
{\tt Reduze}~\cite{Studerus:2009ye, vonManteuffel:2012np}. Cross checks were
obtained using a private version of the program {\tt Crusher} made available
to us by Peter Marquard. For the vertex corrections, analytic results
are presented valid in arbitrary space-time dimensions and for
arbitrary masses. Since for the vertex corrections all master
integrals are known, the leading-colour contribution to the two-loop
amplitude can be calculated numerically, using the results presented
in this article.  For the double-box contributions, we considered a
fixed ratio $\mw^2/\mt^2 = 3/14$, to reduce the complexity of the
calculation. Even for this special case where the number of
independent variables is reduced, the results are too long to be
presented in analytic form. To illustrate the feasibility of the
calculation --- once all the master integrals are known --- we have created a
C library, allowing the calculation of the double-box diagrams. This
library is generated automatically from the analytic results and can
be obtained on demand. 

\section*{Acknowledgments}

We would like to thank Michal Czakon, Thomas Gehrmann and Andrey Grozin
for useful discussions.
We are very grateful to Cedric Studerus and Andreas von Manteuffel for
assistance with {\tt Reduze~2}~\cite{vonManteuffel:2012np}, and to
Peter Marquard for making {\tt Crusher}~\cite{crusher} available to
us. We would also like to thank Sophia Borowka for advice on the use of
{\tt SecDec}~\cite{Borowka:2012yc}, and Guido Bell for sending us an
electronic version of the master integrals of
\Refs{Bell:2006tz,Bell:2007tv,Bell:2009nk}.
The Feynman diagrams in our calculation were generated by
{\tt QGRAF}~\cite{Nogueira:1991ex} and processed further,
in part, using {\tt FORM}~\cite{Vermaseren:2000nd}.
This work is supported by the Helmholtz alliance HA-101 ``Physics at the
Terascale'' and by the Deutsche Forschungsgemeinschaft in the
Sonderforschungs\-bereich/Transregio
SFB/TR-9 ``Computational Particle Physics'' and through the
DFG Research Training Group GRK 1504 ``Mass, Spectrum, Symmetry''.

\appendix
\section{Vertex contribution to $A_{1,SC}^{(2)}$}
\label{sec:A1SCvertex}
The sub-leading colour contribution of the vertex diagrams is given by:
\begin{align}
 (\hat{t} - m_{W}^2) &\times A^{(2)}_{1,SC}=\hspace{-12mm}\notag \\ 
 &\hspace{-12mm}{(\MIOneZeroOnep)^2 \over 128 (-5 + d)^2 (-4 + d)^2 (-3 + d)^2 (-7 + 2 d) 
 (-8 + 3 d) (-1 +  t)^4 t (1 + t) (6 - d - 10 t + 3 d t) } \notag \\
  &\hspace{-12mm}{1 \over (24 - 7 d + 8 t - 2 d t - 4 t^2 + d t^2)}  (d - 2) 
 \notag \\ \times
  \Bigl[ 4 \Shat{1}&  (d - 5)
   \Bigl(
 8 (d - 5) (d - 4)^3 (d - 3) (d - 2) (2 d - 7) (3 d - 10) (3 d - 8) t^9 \notag \\ 
 & +
 (d - 4) (72 d^{10} - 2496 d^9 + 37217 d^8 -  311525 d^7 + 1592803 d^6 - 5001809 d^5 \notag \\ 
 & +
 8776242 d^4 -4594508 d^3 - 12037976 d^2 + 23649600 d -  13143040) t^8 \notag \\ 
 & -
 (d - 4) (456 d^{10} - 15620 d^9 + 234269 d^8 -  2024207 d^7 + 11116257 d^6 - 40224087 d^5 \notag \\ 
 & +
 95689188 d^4 -143362844 d^3 + 120093376 d^2 - 37593696 d -    6961152) t^7 \notag \\ 
 & +
 (360 d^{11} - 13908 d^{10} + 239927 d^9 -  2457445 d^8 + 16681741 d^7 - 78969673 d^6 \notag \\ 
 & +265901534 d^5 -  634846884 d^4 + 1046959112 d^3 - 1125115136 d^2 +  698682368 d \notag \\ 
 & -
 185163776) t^6 \notag \\ 
 & +
 (1944 d^{11} - 73292 d^{10} +   1240123 d^9 - 12446713 d^8 + 82294307 d^7 - 375547289 d^6 \notag \\ 
 & +
 1202264680 d^5 - 2683212260 d^4 + 4050878208 d^3 -    3874266752 d^2 + 2044742144 d \notag \\ 
 & -
 416348160) t^5 \notag \\ 
 & -
 (2472 d^{11} -  92700 d^{10} + 1548463 d^9 - 15221501 d^8 + 97812709 d^7 -  430679297 d^6 \notag \\ 
 & +
 1320775118 d^5 - 2800487420 d^4 +   3968522792 d^3 - 3481789664 d^2 + 1592171520 d \notag \\ 
 & -
 224776192) t^4 \notag \\ 
 & -
 (984 d^{11} - 36524 d^{10} + 643215 d^9 -    7036377 d^8 + 52447151 d^7 - 276033761 d^6 \notag \\ 
 & +
 1035505520 d^5 - 2746965908 d^4 + 5021880176 d^3 - 6000847008 d^2 +   4203434880 d \notag \\ 
 & -
 1302693888) t^3 \notag \\ 
 & +
 (2040 d^{11} - 76092 d^{10} +  1298069 d^9 - 13328407 d^8 + 91238503 d^7 - 435788819 d^6 \notag \\ 
 & +
 1476990730 d^5 - 3539359148 d^4 + 5853548856 d^3 - 6333491392 d^2 + 4010948864 d \notag \\ 
 & -
 1116893184) t^2 \notag \\ 
 & -
 (d -  4) (504 d^{10} - 17308 d^9 + 265247 d^8 - 2379033 d^7 +   13756739 d^6 - 53254969 d^5 \notag \\ 
 & +
 138719412 d^4 - 237663604 d^3 +    252395024 d^2 - 145947456 d + 32799744) t \notag \\ 
 & +
 2 (d - 6) (d - 4) (d - 3)^2 (2 d - 7) (3 d - 8) (7 d -  24) (d^3 - 14 d^2 + 60 d - 84)
     \Bigr) \notag \\ + 
 \Shat{3} &
\Bigl(
4 (d - 4)^2 (d - 3) (d - 2) (2 d - 7) (3 d - 10) (3 d - 8) (d^3 - 
      12 d^2 + 47 d - 64) t^{10} \notag \\ 
 & -
 (d - 4) (39 d^{10} - 758 d^9 + 
      1014 d^8 + 102890 d^7 - 1365961 d^6 + 8953268 d^5  
       \notag \\ 
 & -  35633348 d^4+
 90116280 d^3 - 142168064 d^2 + 128061984 d -   50426624) t^9 \notag \\ 
 & -
   2 (144 d^{12} - 5853 d^{11} + 106233 d^{10} - 1125913 d^9 + 
      7614893 d^8 - 33329432 d^7 \notag \\ 
 & + 
 87548326 d^6- 82676166 d^5 -
      292510688 d^4 + 1319679832 d^3 - 2429628320 d^2  
       \notag \\
 & +
 2295716736 d- 912044032) t^8 \notag \\ 
 & +
   2 (624 d^{12} - 25517 d^{11} + 467011 d^{10} - 5024149 d^9 + 
      34985681 d^8 - 162697988 d^7 \notag \\ 
 & +
 495353856 d^6 - 872953574 d^5 + 
      326362560 d^4 + 2340163592 d^3 - 5851837696 d^2 \notag \\ 
 & +
      6135096256 d - 2561895936) t^7 \notag \\ 
 & +
   2 (240 d^{12} - 7771 d^{11} + 110613 d^{10} - 899449 d^9 \notag \\ 
 & +
 4465165 d^8 - 
      12495678 d^7 + 6198642 d^6 + 106764150 d^5 - 512877072 d^4 \notag \\ 
 & +
      1284522024 d^3 - 1965748896 d^2 + 1745927296 d - 
      694160384) t^6 \notag \\ 
 & -
   2 (2448 d^{12} - 98940 d^{11} + 1793243 d^{10} - 19164393 d^9 + 
      133274773 d^8 - 625172561 d^7 \notag \\ 
 & +1963191740 d^6 - 
      3827692686 d^5 + 3208111312 d^4 + 4073077832 d^3 - 
      14872218784 d^2 \notag \\ 
 & +
 16882063040 d - 7277689344) t^5 \notag \\ 
 & +
   2 (1488 d^{12} - 64421 d^{11} + 1222387 d^{10} - 13385791 d^9 + 
      93186883 d^8 - 424085370 d^7 \notag \\ 
 & + 1218939742 d^6 - 
      1806863846 d^5 - 691010600 d^4 + 9062809384 d^3 - 
      18505002528 d^2 \notag \\ 
 & +
 17734922880 d - 6934335488) t^4 \notag \\ 
 & +
   2 (1488 d^{12} - 58931 d^{11} + 1079209 d^{10} - 12100551 d^9 + 
      92514971 d^8 - 507226248 d^7 \notag \\ 
 & +
  2037633608 d^6 - 
      6011525698 d^5 + 12838263664 d^4 - 19186420904 d^3 
      \notag \\
 &+
 18825711552 d^2 - 10702091712 d + 2585400832) t^3 \notag \\ 
 & -
   2 (1584 d^{12} - 66423 d^{11} + 1264341 d^{10} - 14414743 d^9 + 
      109335375 d^8 - 579143804 d^7 \notag \\ 
 & +
 2185485568 d^6 - 
      5874474730 d^5 + 11022458856 d^4 - 13756521592 d^3 + 
      10301462432 d^2 \notag \\ 
 & -
 3551784064 d + 57738240) t^2 \notag \\ 
 & +
 (672 d^{12} - 
      28945 d^{11} + 560716 d^{10} - 6433612 d^9 + 48420964 d^8 - 
      249708795 d^7 \notag \\ 
 & + 
 892337088 d^6 - 2169762400 d^5 + 
      3360836360 d^4 - 2662256912 d^3 - 284884128 d^2 \notag \\ 
 & +
      2342032896 d - 1353775104) t \notag \\ 
 & -
   4 (d - 6) (d - 5) (d - 4) (d - 3)^2 (2 d - 7) (3 d - 8) (7 d - 
      24) (d^3 - 14 d^2 + 60 d - 84)
 \Bigr)  
 \Bigr]  \notag \\
 &\hspace{-12mm}+{ \MIOneZeroOnep \times \MITwoZeroOne \over 4 (d - 4)^2 (d - 3) (t - 1) } (-2 + d) 
 \times (16 - 7 d + d^2) \notag \\ \times 
 \Bigl[
 \Shat{3}&
(d + t - 3) +
\Shat{1}(d^2 - 9 d + 20) 
 \Bigr]  \notag \\
  &\hspace{-12mm}+{ \MIOneZeroOnep \times \MITwoZeroOnep  \over 32 (d - 5) (d - 4)^2 (d - 3) (2 d - 7) (3 d - 8) (t - 1)^4 (t + 
   1) (d t^2 - 4 t^2 - 2 d t + 8 t - 7 d + 24)} \notag \\ \times 
 \Bigl[
 \Shat{3}&
 \Bigl(
 (d - 4) (d - 2) (3 d - 8) (8 d^6 - 193 d^5 + 1958 d^4 - 10665 d^3 + 
    32936 d^2 - 55008 d \notag \\ 
 & +
 39152) t^7 \notag \\ 
 & +
 2 (24 d^{10} - 877 d^9 + 14654 d^8 - 148051 d^7 + 1003072 d^6 - 
    4755828 d^5 + 15922406 d^4 \notag \\ 
 & -
 36982364 d^3 + 56718880 d^2 - 
    51597952 d + 21047296) t^6 \notag \\ 
 & +
 (-192 d^{10} + 6650 d^9 - 105855 d^8 + 
    1025274 d^7 - 6711491 d^6 + 31010198 d^5 \notag \\ 
 & -
 102014060 d^4 + 
    234445784 d^3 - 357683136 d^2 + 324984832 d - 132818944) t^5 \notag \\ 
 & - 
 4 (d - 4) (36 d^9 - 1143 d^8 + 15875 d^7 - 129159 d^6 + 689474 d^5 - 
    2531265 d^4 \notag \\ 
 & +
 6408378 d^3 - 10727464 d^2 + 10654464 d - 
    4723456) t^4 \notag \\ 
 & +
 (768 d^{10} - 27668 d^9 + 457159 d^8 - 4565336 d^7 + 
    30487811 d^6 - 141920362 d^5 \notag \\ 
 & +
 464681264 d^4 - 1052303520 d^3 + 
    1570727840 d^2 - 1390172032 d + 552221696) t^3 \notag \\ 
 & - 
 2 (120 d^{10} - 4895 d^9 + 93758 d^8 - 1074277 d^7 + 8014432 d^6 - 
    40380396 d^5 + 138851282 d^4 \notag \\ 
 & -
 321810020 d^3 + 481706112 d^2 - 
    421213824 d + 163678208) t^2 \notag \\ 
 & +
 (-576 d^{10} + 19938 d^9 - 
    310357 d^8 + 2885938 d^7 - 17897281 d^6 + 77838658 d^5 \notag \\ 
 & -
    241230924 d^4 + 525745464 d^3 - 768104064 d^2 + 675136512 d - 
    269293568) t \notag \\ 
 & +
 4 (d - 4) (2 d - 7) (3 d - 8) (7 d - 24) (2 d^6 - 45 d^5 + 440 d^4 - 
    2400 d^3 + 7663 d^2 \notag \\ 
 & -
 13416 d + 9916)
 \Bigr) \notag \\ 
+ 4 \Shat{1}& (d-4 )
 \Bigl(
 (d - 4) (d - 2) (3 d - 8) (16 d^5 - 392 d^4 + 3733 d^3 - 17410 d^2 + 
    39947 d - 36170) t^6 \notag \\ 
 & +
 (24 d^9 - 1300 d^8 + 24690 d^7 - 
    249483 d^6 + 1552630 d^5 - 6306395 d^4 + 16879494 d^3 \notag \\ 
 & -
    28802184 d^2 + 28423376 d - 12335232) t^5 \notag \\ 
 & - 
 2 (36 d^9 - 1398 d^8 + 21038 d^7 - 170071 d^6 + 835325 d^5 - 
    2632780 d^4 + 5431774 d^3 \notag \\ 
 & -
 7272136 d^2 + 5948248 d - 
    2346816) t^4 \notag \\ 
 & -
 2 (72 d^9 - 4140 d^8 + 76608 d^7 - 734679 d^6 + 4286438 d^5 - 
    16242095 d^4 + 40585888 d^3 \notag \\ 
 & -
 64992352 d^2 + 60688024 d - 
    25162304) t^3 \notag \\ 
 & +
 (240 d^9 - 8152 d^8 + 122076 d^7 - 1068369 d^6 + 
    6055692 d^5 - 23106495 d^4 + 59338384 d^3 \notag \\ 
 & -
 98712036 d^2 + 
    96264800 d - 41807360) t^2 \notag \\ 
 & +
 (120 d^9 - 6980 d^8 + 126510 d^7 - 
    1171107 d^6 + 6523462 d^5 - 23410323 d^4 + 55174090 d^3 \notag \\ 
 & -
    83395432 d^2 + 73915776 d - 29360896) t \notag \\ 
 & -
 2 (2 d - 7) (3 d - 8) (7 d - 24) (2 d^6 - 44 d^5 + 424 d^4 - 
    2301 d^3 + 7367 d^2 - 12988 d + 9676)
 \Bigr)  
\Bigr]  \notag \\
  &\hspace{-12mm}-{ (\MITwoZeroOne)^{2}  \over 4 (d - 4)^2} \times \Bigl[
 \Shat{3}  (d^2 - 7 d + 16)^2
 \Bigr] \notag \\
  &\hspace{-12mm}+{ \MITwoZeroOne \times \MITwoZeroOnep \over 4 (d - 4)^2 (t - 1)} (d^2 - 7 d + 16) \notag \\ \times 
 \Bigl[
 \Shat{3} &
 \Bigl(
 (d - 5) (d - 4) - (d^2 - 7 d + 16) t
 \Bigr) -
 \Shat{1}
  \Bigl(
 2 (d - 5) (d - 4) 
 \Bigr)
  \Bigr]  \notag \\
  &\hspace{-12mm}-{  \text{(MI201p)}^2 \over 4 (d - 4)^2 (d - 2) (t - 1)^2}\notag \\ \times 
 \Bigl[
   \Shat{1}&
  \Bigl(
  2 (d - 5) (d - 4) (d - 2) (d^2 - 7 d + 20) t -  2 (d - 5) (d - 4)^3 (d - 3)
 \Bigr) \notag \\ 
+  \Shat{3} & 
 \Bigl(
 (d - 5) (d - 3) (d - 4)^3 -   2 (d - 5) (d^3 - 9 d^2 + 31 d - 36) t (d - 4) \notag \\ 
 & +
 (d - 2) (d^2 -   7 d + 16)^2 t^2
 \Bigr)
\Bigr]  \notag \\
   &\hspace{-12mm}-{  \MIThreeZeroOne \over 16 (d - 4)^3 (d - 3) (2 d - 7) t}\notag \\ \times 
 \Bigl[
 \Shat{3}&
 \Bigl(
 (3 d - 8) (9 d^6 - 358 d^5 + 4309 d^4 - 24466 d^3 + 72896 d^2 - 
   110064 d + 66080)
   \Bigr)
 \Bigr]  \notag \\
   &\hspace{-12mm}+{  \MIThreeZeroOnep  \over  384 (d - 5) (d - 4)^2 (d - 3) (d - 2) (2 d - 7) (3 d - 14) (3 d - 
   10) (t - 1)^4 t } \notag \\
 &\hspace{-12mm}{ 1 \over (t + 1) (3 t d - d - 10 t + 6) } (3 d - 8) \notag \\ \times 
 \Bigl[
 4 \Shat{1 } &(d - 5) (d -  4) 
 \Bigl(-(d - 2) (3 d - 10) (504 d^6 - 13635 d^5 + 150321 d^4 - 
       867796 d^3 \notag \\ 
 & +
 2774132 d^2 - 4665168 d + 
       3229056) t^6 \notag \\ 
 & +
 (20160 d^8 - 540981 d^7 + 6439113 d^6 - 
       44379614 d^5 + 193384052 d^4 \notag \\ 
 & -
 544063128 d^3 + 961795520 d^2 - 
       973087872 d + 429685632) t^5 \notag \\ 
 & -
    2 (11700 d^8 - 310191 d^7 + 3684359 d^6 - 25590326 d^5 + 
       113383732 d^4 - 326829192 d^3 \notag \\ 
 & +
 595654144 d^2 - 624465984 d + 
       286989312) t^4 \notag \\ 
 & -
    2 (14688 d^8 - 341151 d^7 + 3411009 d^6 - 19076404 d^5 + 
       64722548 d^4 - 134442944 d^3 \notag \\ 
 & +
 162294832 d^2 - 97231104 d + 
       17141184) t^3 \notag \\ 
 & +
 (14472 d^8 - 407187 d^7 + 5005133 d^6 - 
       35204948 d^5 + 155353288 d^4 - 441131248 d^3 \notag \\ 
 & +
 787549200 d^2 - 
       807657024 d + 363577344) t^2 \notag \\
 &
 - (1152 d^8 - 42423 d^7 + 
       673815 d^6 - 6011238 d^5 + 32891740 d^4 - 112997480 d^3 \notag \\ 
 & +
       238155264 d^2 - 281853504 d + 143604864) t \notag \\ 
 & +
    4 (10 - 3 d)^2 (d - 6) (d - 3) (2 d - 7) (d^3 - 14 d^2 + 60 d -  84)
       \Bigr)  \notag \\ +
\Shat{3} &
\Bigl(
-(10 - 3 d)^2 (d - 2) (129 d^7 - 3212 d^6 +   34391 d^5 - 205404 d^4 + 
739512 d^3 - 1607360 d^2 \notag \\ 
 & +
 1961376 d - 1046016) t^7 \notag \\ 
 & +
    4 (3 d - 10) (849 d^9 - 30032 d^8 + 473927 d^7 - 4373749 d^6 + 
       25975613 d^5 - 102786064 d^4 \notag \\ 
 & +
 270537788 d^3 - 455956880 d^2 + 
       445763712 d - 192286080) t^6 \notag \\ 
 & -
 (25209 d^{10} - 1033338 d^9 + 
       19009043 d^8 - 206384706 d^7 + 1463111000 d^6 - 
       7071516464 d^5 \notag \\ 
 & +
 23585007664 d^4 - 53575590816 d^3 + 
       79301446912 d^2 - 69045440512 d + 26844840960) t^5 \notag \\ 
 & +
    8 (2475 d^{10} - 113604 d^9 + 2251836 d^8 - 25692899 d^7 + 
       188222312 d^6 - 929348420 d^5 \notag \\ 
 & +
 3141614008 d^4 - 
       7194499824 d^3 + 10697199936 d^2 - 9334245056 d + 
       3632292480) t^4 \notag \\ 
 & +
 (4077 d^{10} - 75930 d^9 + 212327 d^8 + 
       6377826 d^7 - 84114572 d^6 + 512009736 d^5 \notag \\ 
 & -
 1854381088 d^4 + 
       4202905216 d^3 - 5839036544 d^2 + 4524404480 d - 
       1481917440) t^3 \notag \\ 
 & -
    4 (2709 d^{10} - 111876 d^9 + 2080489 d^8 - 22923137 d^7 + 
       165559865 d^6 - 818174674 d^5 \notag \\ 
 & +
 2799104716 d^4 - 
       6540192632 d^3 + 9980180672 d^2 - 8975227072 d + 
       3609980160) t^2 \notag \\ 
 & +
 (3285 d^{10} - 146778 d^9 + 2917887 d^8 -  33985466 d^7 + 256862728 d^6 -
 1316665488 d^5 \notag \\ 
 & +
 4637311856 d^4 - 11085869472 d^3 + 17223742720 d^2 - 
 15713037312 d + 6395443200) t \notag \\ 
 & -
 8 (10 - 3 d)^2 (d - 6) (d - 5) (d - 4) (d - 3) (2 d - 7) (d^3 -  14 d^2 + 60 d - 84)
       \Bigr)
 \Bigr]  \notag \\
   &\hspace{-12mm}+{  \MIThreeZeroOnepuOne \over   256 (d - 5) (d - 4)^2 (d - 3) (d - 2) (2 d - 7)^2 (t - 
   1)^3 t (d t^2 - 4 t^2 - 2 d t + 8 t - 7 d + 24)}\notag \\ \times 
 \Bigl[
4 \Shat{1} & (d - 5) (d -  4) 
 \Bigl(
 -4 (d - 4) (d - 2) (50 d^6 - 1311 d^5 + 13537 d^4 -  71690 d^3 \notag \\ 
 & +
 207408 d^2 - 312720 d + 192736) t^5 \notag \\ 
 & +
 (2552 d^8 -    69101 d^7 + 822780 d^6 - 5613873 d^5 + 23949730 d^4 -    65265572 d^3 \notag \\ 
 & +
 110703544 d^2 - 106645280 d +  44591360) t^4 \notag \\ 
 & -
 (1072 d^8 - 27329 d^7 + 330514 d^6 -    2440765 d^5 + 11734452 d^4 - 36715452 d^3 \notag \\ 
 & +
 71785904 d^2 -   79396032 d + 37838336) t^3 \notag \\ 
 & -
 (16496 d^8 - 400017 d^7 +     4278992 d^6 - 26361489 d^5 + 102197258 d^4 \notag \\ 
 & -
 254892476 d^3 +  398579064 d^2 - 356380672 d + 139113984) t^2 \notag \\ 
 & -
 (10376 d^8 -   214955 d^7 + 1847346 d^6 - 8281675 d^5 + 19121020 d^4 - 
 13469588 d^3 \notag \\ 
 & -
 34656880 d^2 + 82573312 d - 53170176) t \notag \\ 
 & +
 4 (d - 3) (2 d - 7) (3 d - 10) (3 d - 8) (7 d - 24) (d^3 -   14 d^2 + 60 d - 84)
 \Bigr) \notag \\ +
  \Shat{3} &  
  \Bigl(
  -8 (d - 4) (d - 2) (3 d - 8) (3 d^7 - 119 d^6 + 1743 d^5 - 
    13193 d^4 + 57458 d^3 - 146044 d^2 \notag \\ 
 & +
 201960 d -     117560) t^6 \notag \\ 
 & +
(d - 3) (1749 d^9 - 61760 d^8 + 958255 d^7 - 
8593272 d^6 + 49142988 d^5 - 185948688 d^4 \notag \\ 
 & +
 465510048 d^3 - 743169984 d^2 + 686097664 d - 278865920) t^5 \notag \\ 
 & -
4 (1425 d^{10} - 53383 d^9 + 900152 d^8 - 8990014 d^7 + 58847091 d^6 - 
263613479 d^5 \notag \\ 
 & +
 817847588 d^4 - 1733966660 d^3 + 2402720288 d^2 -  1963651008 d + 718318080) t^4 \notag \\ 
 & -
2 (525 d^{10} - 11849 d^9 + 80259 d^8 + 263195 d^7 - 7904088 d^6 + 
60076606 d^5 - 255495856 d^4 \notag \\ 
 & +
 676725336 d^3 - 1114395648 d^2 + 1050515200 d - 434903040) t^3 \notag \\ 
 & +
4 (2889 d^{10} - 112075 d^9 + 1935886 d^8 - 19637010 d^7 + 
129674329 d^6 - 582890855 d^5 \notag \\ 
 & +
 1807046696 d^4 - 3816289204 d^3 + 5255585008 d^2 - 4262445824 d + 1546163200) t^2 \notag \\ 
 & +
(-5475 d^{10} +   235357 d^9 - 4378649 d^8 + 46943563 d^7 - 323389276 d^6 + 
1502599664 d^5 \notag \\ 
 & -
 4784211872 d^4 + 10330949008 d^3 - 14504696576 d^2 + 11972023296 d - 4415692800) t \notag \\ 
 & -
8 (d - 5) (d - 4) (d - 3) (2 d - 7) (3 d - 10) (3 d - 8) (7 d - 
24) (d^3 - 14 d^2 + 60 d - 84)
   \Bigr) 
\Bigr]   \notag \\
   &\hspace{-12mm}+{  \MIThreeZeroTwop  \over 384 (d - 5) (d - 4)^2 (d - 3)^2 (d - 2) (2 d - 7) (3 d - 14) (3 d - 
   10) (t - 1)^4 t  } \notag \\
   &\hspace{-12mm}{  1  \over  (t + 1) (3 t d - d - 10 t + 6) }\notag \\ \times 
 \Bigl[
4 \Shat{1}   & (d - 5) (d - 4) 
 \Bigl(
 -(d - 4) (d - 2) (3 d - 10) (504 d^6 - 11637 d^5 + 113655 d^4 -   
 604084 d^3 \notag \\ 
 & +
 1847396 d^2 - 3079344 d + 2177376) t^7 \notag \\ 
 & +
 2 (11916 d^9 - 414225 d^8 + 6352830 d^7 - 56374909 d^6 + 
    318836394 d^5 - 1191362572 d^4 \notag \\ 
 & +
 2940103576 d^3 - 
    4619041056 d^2 + 4189615488 d - 1670289024) t^6 \notag \\ 
 & -
 (103464 d^9 - 
    2938629 d^8 + 37503853 d^7 - 283016178 d^6 + 1394564364 d^5 - 
    4658982648 d^4 \notag \\ 
 & +
 10555759136 d^3 - 15629324544 d^2 + 
    13702223232 d - 5408057088) t^5 \notag \\ 
 & -
 4 (29610 d^9 - 782304 d^8 + 9109423 d^7 - 61230473 d^6 + 
    260959440 d^5 - 727545024 d^4 \notag \\ 
 & +
 1315807272 d^3 - 1467083328 d^2 +  889375008 d - 208577472) t^4 \notag \\ 
 & +
 (23112 d^9 - 886839 d^8 + 
    14132645 d^7 - 126227280 d^6 + 707027656 d^5 - 2599976032 d^4 \notag \\ 
 & +
    6316590928 d^3 - 9818071936 d^2 + 8883495168 d - 
    3570219264) t^3 \notag \\ 
 & +
 2 (5796 d^9 - 154869 d^8 + 1736686 d^7 - 10372501 d^6 + 
    33208662 d^5 - 38047564 d^4 \notag \\ 
 & -
 102311352 d^3 + 454295424 d^2 - 
    675620352 d + 378190080) t^2 \notag \\ 
 & -
 (1080 d^9 - 39603 d^8 + 
    647143 d^7 - 6166938 d^6 + 37643332 d^5 - 152141288 d^4 \notag \\ 
 & +
    405964928 d^3 - 687895680 d^2 + 670167552 d - 285405696) t \notag \\ 
 & +
 4 (10 - 3 d)^2 (d - 6) (d - 4) (d - 3) (2 d - 7) (d^3 - 14 d^2 + 
    60 d - 84)
 \Bigr)    \notag \\ 
+  \Shat{3} &
  \Bigl(
  -(10 - 3 d)^2 (d - 4) (d - 2) (111 d^7 - 2300 d^6 + 19457 d^5 - 
    85596 d^4 + 203904 d^3 \notag \\ 
 & -
 241472 d^2 + 101664 d -  384) t^8 + (3 d - 10) (5385 d^{10} - 226808 d^9 + 4182179 d^8 \notag \\ 
 & -
    44748716 d^7 + 309005984 d^6 - 1443157120 d^5 + 4626129232 d^4 - 
    10065576512 d^3 \notag \\ 
 & +
 14242561536 d^2 - 11844667392 d +     4399303680) t^7 \notag \\ 
 & +
 (-55899 d^{11} + 2430714 d^{10} - 48230193 d^9 + 
    575086382 d^8 - 4570303756 d^7 \notag \\ 
 & +
 25383400200 d^6 -     100436950272 d^5 + 282934787520 d^4 - 555873116288 d^3 \notag \\ 
 & +
    725240141568 d^2 - 565491947008 d +    199649648640) t^6 \notag \\ 
 & +
 (45999 d^{11} - 2250018 d^{10} + 48382693 d^9 - 
    609765946 d^8 + 5035567328 d^7 \notag \\ 
 & -
 28727760016 d^6 + 
    115852614736 d^5 - 330909888800 d^4 + 656973248128 d^3 \notag \\ 
 & -
    864318028288 d^2 + 678672041984 d - 
    241094123520) t^5 \notag \\ 
 & +
 (20475 d^{11} - 620298 d^{10} + 8041961 d^9 - 
    56735830 d^8 + 219514604 d^7 \notag \\ 
 & -
 301492744 d^6 - 1261549408 d^5 + 
    7829491520 d^4 - 19703007104 d^3 \notag \\ 
 & +
 27190314752 d^2 - 
    19850728448 d + 5852774400) t^4 \notag \\ 
 & +
 (-27279 d^{11} + 1276794 d^{10} - 
    26458565 d^9 + 323448330 d^8 - 2607788888 d^7 \notag \\ 
 & +
 14621602688 d^6 - 
    58346744400 d^5 + 166033462048 d^4 - 330601074432 d^3 \notag \\ 
 & +
    438997027072 d^2 - 349978580992 d + 
    126904350720) t^3 \notag \\ 
 & +
 (-1449 d^{11} + 846 d^{10} + 1108325 d^9 - 
    25258278 d^8 + 286327628 d^7 - 2018882504 d^6 \notag \\ 
 & +
 9546729632 d^5 - 
    31042513536 d^4 + 68958355968 d^3 - 100505389568 d^2 \notag \\ 
 & +
    86971736576 d - 33976074240) t^2 \notag \\ 
 & +
 (d - 4) (3141 d^{10} - 
    131202 d^9 + 2450879 d^8 - 26942706 d^7 + 192880408 d^6 \notag \\ 
 & -
    938885584 d^5 + 3144719824 d^4 - 7150909152 d^3 + 
    10556251264 d^2 - 9126496512 d \notag \\ 
 & +
 3505305600) t \notag \\ 
 & -
 8 (10 - 3 d)^2 (d - 6) (d - 5) (d - 4)^2 (d - 3) (2 d - 7) (d^3 - 
    14 d^2 + 60 d - 84)
  \Bigr)
\Bigr]   \notag \\
   &\hspace{-12mm}+{  \MIThreeZeroThreep  \over  8 (d - 4)^2 (d - 3) (d - 2) (t - 1)^3 (t + 1)} \notag \\ \times 
 \Bigl[
  \Shat{3} & 
  \Bigl(
  (d - 2) (3 d - 8) (d^3 - 10 d^2 + 33 d - 37) t^4 \notag \\ 
 & +(-13 d^5 + 
    230 d^4 - 1605 d^3 + 5556 d^2 - 9568 d + 6560) t^3 \notag \\ 
 & +(17 d^5 - 
    216 d^4 + 1003 d^3 - 1946 d^2 + 1052 d + 640) t^2 \notag \\ 
 & +(-23 d^5 + 
    310 d^4 - 1595 d^3 + 3804 d^2 - 3936 d + 1120) t \notag \\ 
 & + (2 d - 
    7) (8 d^4 - 112 d^3 + 579 d^2 - 1314 d + 1104)
  \Bigr) \notag \\ 
+   2 \Shat{1} &
  \Bigl(
  (d - 2) (d^5 - 7 d^4 - 39 d^3 + 483 d^2 - 1498 d + 
    1520) t^3 \notag \\ 
 & + (-5 d^6 + 117 d^5 - 1079 d^4 + 5059 d^3 - 12712 d^2 + 
    16128 d - 7968) t^2 \notag \\ 
 & + (-9 d^6 + 249 d^5 - 2367 d^4 + 10759 d^3 - 
    25280 d^2 + 29124 d - 12576) t \notag \\ 
 & -
 (d - 4) (3 d^5 - 39 d^4 +   149 d^3 - 57 d^2 - 684 d + 968)
  \Bigr)
 \Bigr]   \notag \\
   &\hspace{-12mm}+{  \MIFourZeroOne  \over  16 (d - 4)^2 (2 d - 7) (3 d - 8)}\notag \\ \times 
 \Bigl[
 \Shat{3} &
  \Bigl(
 21 d^6 - 789 d^5 + 9422 d^4 - 53864 d^3 + 163200 d^2 - 
 253472 d + 159232
  \Bigr)
\Bigr]  \notag \\
   &\hspace{-12mm}+{  \MIFourZeroTwop  \over  32 (d - 4)^2 (d - 2) (2 d - 7) (3 d - 8) (t - 1)^3 (t + 1) } \notag \\ \times 
 \Bigl[
4  \Shat{1} & (d - 4)  
 \Bigl(
 -(d - 2) (4 d^5 + 145 d^4 - 2501 d^3 + 13664 d^2 - 31732 d + 
    27040) t^3 \notag \\ 
 & -
 (140 d^6 - 2243 d^5 + 16659 d^4 - 72088 d^3 + 
    184420 d^2 - 255344 d + 146176) t^2 \notag \\ 
 & -
 2 (2 d - 7) (35 d^5 - 298 d^4 + 632 d^3 + 1196 d^2 - 6008 d + 
    5888) t \notag \\ 
 & -
 2 (2 d - 7) (d^5 - 20 d^4 + 176 d^3 - 788 d^2 + 1704 d - 1408)
 \Bigr) \notag \\ 
+  \Shat{3} &
  \Bigl(
  (d - 2) (3 d^6 - 475 d^5 + 7498 d^4 - 49896 d^3 + 168688 d^2 - 
    286624 d + 194816) t^4 \notag \\ 
 & -
 2 (39 d^7 - 1375 d^6 + 18274 d^5 - 125446 d^4 + 491184 d^3 - 
    1109048 d^2 \notag \\ 
 & +
 1345440 d - 679168) t^3 \notag \\ 
 & +(3 d - 8) (57 d^6 - 
    1575 d^5 + 16796 d^4 - 90904 d^3 + 266832 d^2 - 405520 d + 
    250112) t^2 \notag \\ 
 & -
 4 (2 d - 7) (15 d^6 - 380 d^5 + 3722 d^4 - 18616 d^3 + 50984 d^2 - 
    73088 d + 43008) t \notag \\ 
 & +
 4 (d - 4) (2 d - 7) (3 d^5 - 46 d^4 + 304 d^3 - 1084 d^2 + 2024 d -  1536)
  \Bigr) 
 \Bigr]  \notag \\
   &\hspace{-12mm}+{  \text{MI403p}  \over   64 (d - 5) (d - 4) (d - 3) (d - 2) (2 d - 7) (t - 1)^3 (-10 t + 
   d (3 t - 1) + 6)} \notag \\ \times 
 \Bigl[
\Shat{1}&
\Bigl(
4 (d - 5) (d - 4) (d - 2)^2 (3 d - 10) (37 d^2 - 309 d + 626) t^4 \notag \\ 
 & -
 8 (d - 5) (d - 4) (d - 2) (3 d - 10) (40 d^3 - 495 d^2 + 1997 d - 
    2646) t^3 \notag \\ 
 & +
 8 (d - 5) (347 d^6 - 6556 d^5 + 51757 d^4 - 218326 d^3 + 
    518564 d^2 - 657064 d + 346720) t^2 \notag \\ 
 & -
 8 (d - 5) (104 d^6 - 2031 d^5 + 16285 d^4 - 69386 d^3 + 167380 d^2 - 
    218536 d + 121184) t \notag \\ 
 & -
 4 (d - 6) (d - 5) (d - 4) (d - 2) (37 d^3 - 365 d^2 + 1070 d - 872)
\Bigr) \notag \\ + 
 \Shat{3} &
\Bigl(
-(d - 2)^2 (3 d - 10) (d^5 - 44 d^4 + 527 d^3 - 2732 d^2 + 6624 d - 
    6224) t^5 \notag \\ 
 & +
 (d - 2) (3 d - 10) (27 d^6 - 790 d^5 + 9269 d^4 - 
    56262 d^3 + 187692 d^2 - 327848 d \notag \\ 
 & +
 234880) t^4 \notag \\ 
 & -
 2 (3 d - 10) (31 d^7 - 950 d^6 + 12125 d^5 - 83610 d^4 + 
    337148 d^3 - 796432 d^2 \notag \\ 
 & +
 1021712 d - 549440) t^3 \notag \\ 
 & + 
 2 (67 d^8 - 2284 d^7 + 33437 d^6 - 274288 d^5 + 1381244 d^4 - 
    4383096 d^3 \notag \\ 
 & +
 8580512 d^2 - 9494560 d + 4553600) t^2 \notag \\ 
 & +
 (-19 d^8 + 
    614 d^7 - 8801 d^6 + 72650 d^5 - 379420 d^4 + 1290968 d^3 - 
    2797600 d^2 \notag \\ 
 & +
 3511168 d - 1933440) t \notag \\ 
 & -
 (d - 6) (d - 2) (7 d^6 -  226 d^5 + 2817 d^4 - 17290 d^3 + 55364 d^2 - 87592 d + 52800)
\Bigr)
\Bigr]  \notag \\
   &\hspace{-12mm}+{  \text{MI404p}  \over  8 (d - 4)^2 (d - 3) (d - 2) (2 d - 7) (t - 1)^3 
   (3 t d - d - 10 t + 6)  }  t^2 \notag \\ \times 
 \Bigl[
4 \Shat{1} & (d - 5) (d - 4) 
\Bigl(
-(d - 4) (d - 2) (37 d^2 - 309 d + 626) t^2 \notag \\ 
 & -
4 (4 d^4 - 42 d^3 + 108 d^2 + 71 d - 374) t \notag \\ 
 & + (d - 2) (37 d^3 - 365 d^2 + 1070 d - 872)
\Bigr)\notag \\
 + \Shat{3}&
\Bigl(
(d - 4) (d - 2) (d^5 - 44 d^4 + 527 d^3 - 2732 d^2 + 6624 d - 
    6224) t^3 \notag \\ 
 & +
 (5 d^7 - 140 d^6 + 1255 d^5 - 3688 d^4 - 8584 d^3 + 
    83112 d^2 - 198272 d + 163392) t^2 \notag \\ 
 & +
 (-13 d^7 + 430 d^6 - 
    5323 d^5 + 32858 d^4 - 108592 d^3 + 183032 d^2 - 120048 d - 
    8000) t \notag \\ 
 & +(d - 2) (7 d^6 - 226 d^5 + 2817 d^4 - 17290 d^3 + 
    55364 d^2 - 87592 d + 52800)
\Bigr)
\Bigr]   \notag \\
   &\hspace{-12mm}+{  \MIFourZeroSevenp  \over  8 (d - 5) (d - 4) (d - 3) (d - 2) (3 d - 8) (t - 1)^3}
   \times  \notag \\
\Bigl[
8 \Shat{1} & (d - 5)
\Bigl(
-(d - 5) (d - 4) (d - 2) (3 d - 10) (3 d -   8) t^3 \notag \\ 
 & +
(d - 4) (d^5 - 13 d^4 + 42 d^3 + 46 d^2 - 424 d +  528) t^2 \notag \\ 
 & +
(6 d^6 - 167 d^5 + 1653 d^4 - 7988 d^3 + 20388 d^2 -   26312 d + 13440) t \notag \\ 
 & +
(d - 4) (d^5 - 7 d^4 - 40 d^3 + 450 d^2 -   1272 d + 1168)
\Bigr)  \notag \\+
\Shat{3} &
\Bigl(
6 d^7 - 203 d^6 + 2774 d^5 - 20145 d^4 + 84744 d^3 - 207740 d^2 + 
275904 d - 153600\notag \\ 
 & -
 (d - 2) (3 d - 10) (3 d - 8) (d^3 - 12 d^2 + 47 d -   64) t^4 \notag \\ 
 & -
2 (3 d^7 - 103 d^6 + 1415 d^5 - 10273 d^4 + 43102 d^3 -   105320 d^2 + 139456 d - 77440) t^3 \notag \\ 
 & +
2 (d - 2) (9 d^6 - 260 d^5 + 2955 d^4 - 17152 d^3 + 54140 d^2 -     88592 d + 58880) t^2 \notag \\ 
 & -
2 (d - 2) (9 d^6 - 263 d^5 + 3011 d^4 - 17577 d^3 + 55748 d^2 -    91608 d + 61120) t 
\Bigr)
\Bigr]   \notag \\
   &\hspace{-12mm}+{  \MIFourZeroEightp  \over 4 (d - 5) (d - 4) (d - 3) (d - 2) (3 d - 8) (t - 1) (t + 1) } \notag \\ \times 
\Bigl[
   \Shat{3} &
 \Bigl(
 -(d - 2) (3 d - 8) (d^3 - 12 d^2 + 47 d - 64) t^4 \notag \\ 
 & +
 8 (d - 5) (2 d^4 - 28 d^3 + 144 d^2 - 325 d + 272) t^3 \notag \\ 
 & -
 2 (13 d^5 - 198 d^4 + 1139 d^3 - 3018 d^2 + 3500 d - 1216) t^2 \notag \\ 
 & +
 8 (d - 5) (2 d - 7) (2 d^3 - 14 d^2 + 29 d - 16) t \notag \\ 
 & -
 (d - 4) (19 d^4 - 278 d^3 + 1485 d^2 - 3462 d + 2976)
 \Bigr) \notag \\ 
+  \Shat{1}& 
 \Bigl(
-8 (d - 5)^2 (d - 4) (d - 2) (3 d - 8) t^3 \notag \\ 
 & +
 8 (d - 5) (d^5 - 19 d^4 + 136 d^3 - 456 d^2 + 702 d - 384) t^2 \notag \\ 
 & +
 8 (d - 5) (2 d^5 - 53 d^4 + 405 d^3 - 1322 d^2 + 1908 d - 960) t \notag \\ 
 & +
 8 (d - 5) (d^5 - 13 d^4 + 54 d^3 - 52 d^2 - 146 d + 256)
\Bigr)
   \Bigr]   \notag \\
   &\hspace{-12mm}+{  \text{MI409p}  \over 16 (d - 4) (d - 2) (2 d - 7) (3 d - 8) (t - 1)^3 (d t^2 - 4 t^2 - 
   2 d t + 8 t - 7 d + 24)} \notag \\ \times 
   \Bigl[
   4 \Shat{1}& (d - 4) 
   \Bigl(
   -32 d^6 - 218 d^5 + 10566 d^4 - 89360 d^3 + 
329976 d^2 -   573216 d + 383104 \notag \\ 
 & -
(d - 2) (8 d^5 - 147 d^4 + 1083 d^3 - 3940 d^2 + 
7060 d - 5024) t^4 \notag \\ 
 & +
 (d - 2) (3 d - 8) (24 d^4 - 427 d^3 + 
2681 d^2 - 7074 d + 6616) t^3 \notag \\ 
 & +
 (3 d - 8) (24 d^5 - 419 d^4 + 
2717 d^3 - 8314 d^2 + 12060 d - 6568) t^2 \notag \\ 
 & -
(360 d^6 - 8561 d^5 + 79833 d^4 - 379664 d^3 + 979684 d^2 - 1307264 d +   707072) t 
\Bigr) \notag \\ + 
 \Shat{3} & 
\Bigl(
(d - 2) (21 d^6 - 399 d^5 + 3134 d^4 - 12848 d^3 + 28576 d^2 - 
    32032 d + 13568) t^5 \notag \\ 
 & -
 2 (d - 2) (39 d^6 - 673 d^5 + 4556 d^4 - 14898 d^3 + 22580 d^2 - 
    9648 d - 6016) t^4 \notag \\ 
 & +
 2 (3 d - 8) (19 d^6 - 327 d^5 + 2082 d^4 - 5798 d^3 + 5144 d^2 + 
    5592 d - 9632) t^3 \notag \\ 
 & -
 4 (93 d^7 - 2621 d^6 + 29943 d^5 - 183067 d^4 + 652884 d^3 - 
    1364764 d^2 + 1552016 d \notag \\ 
 & -
 741504) t^2 \notag \\ 
 & +
 (609 d^7 - 18485 d^6 +  226344 d^5 - 1477312 d^4 + 5605424 d^3 - 12432304 d^2 \notag \\ 
 & +
    14971904 d - 7567360) t \notag \\ 
 & -
 2 (147 d^7 - 4603 d^6 + 58212 d^5 - 392120 d^4 + 1533520 d^3 - 
    3500784 d^2 \notag \\ 
 & +
 4333824 d - 2249216)
\Bigr)
\Bigr]  \notag \\
   &\hspace{-12mm}+{  \text{MI410p}  \over 8 (d - 4)^2 (d - 2) (2 d - 7) (3 d - 8) (t - 1)^2 (d t^2 - 4 t^2 - 
   2 d t + 8 t - 7 d + 24)} \notag \\ \times 
 \Bigl[
 4 \Shat{1} & (d - 4)  
 \Bigl(
 (d - 4) (d - 3) (d - 2) (3 d - 8) (5 d - 22) t^4 \notag \\ 
 & +
 (d - 2) (24 d^5 - 599 d^4 + 5303 d^3 - 21846 d^2 + 42692 d -   31904) t^3 \notag \\ 
 & -
 (48 d^6 - 1255 d^5 + 12775 d^4 - 66022 d^3 +    184160 d^2 - 263888 d + 152192) t^2 \notag \\ 
 & -
 (24 d^6 - 515 d^5 +   3275 d^4 - 4922 d^3 - 21408 d^2 + 80800 d - 74624) t \notag \\ 
 & +
 2 (24 d^6 - 557 d^5 + 4749 d^4 - 19318 d^3 + 38976 d^2 - 34784 d +    8320)
 \Bigr) \notag \\ 
+  \Shat{3} &
 \Bigl(
 -(d - 4) (d - 3) (d - 2) (3 d - 8) (d^3 - 20 d^2 + 104 d - 176) t^5 \notag \\ 
 & +
 2 (d - 2) (24 d^6 - 595 d^5 + 5945 d^4 - 30866 d^3 + 88216 d^2 - 
    131856 d + 80512) t^4 \notag \\ 
 & -
 4 (45 d^7 - 1158 d^6 + 12587 d^5 - 74941 d^4 + 264056 d^3 - 
    550572 d^2 + 628624 d \notag \\ 
 & -
 302976) t^3 \notag \\ 
 & + 
 2 (135 d^7 - 3426 d^6 + 37341 d^5 - 226400 d^4 + 823376 d^3 - 
    1791528 d^2 \notag \\ 
 & +
 2152128 d - 1097216) t^2 \notag \\ 
 & +
 (-177 d^7 + 4477 d^6 - 
    49638 d^5 + 311844 d^4 - 1191016 d^3 + 2741632 d^2 \notag \\ 
 & -
 3491328 d + 
    1883136) t \notag \\ 
 & +
 2 (d - 4) (21 d^6 - 449 d^5 + 4302 d^4 - 23164 d^3 + 71416 d^2 - 
    116160 d + 76544)
 \Bigr)
 \Bigr]  \notag \\
   &\hspace{-12mm}+{  \text{MI411p}  \over   16 (d - 4) (d - 3) (d - 2) (2 d - 7) (3 d - 8) (t - 1) (d t^2 - 
   4 t^2 - 2 d t + 8 t - 7 d + 24) } \notag \\ \times 
 \Bigl[
 \Shat{1} &
 \Bigl(
 -4 (d - 4)^2 (d - 3) (d - 2) (3 d - 8) (5 d - 22) t^3 \notag \\ 
 & +
 8 (d - 4) (d - 2) (8 d^5 - 155 d^4 + 1187 d^3 - 4530 d^2 +   8622 d - 6512) t^2 \notag \\ 
 & -
 4 (d - 4) (32 d^6 - 951 d^5 + 10537 d^4 - 58056 d^3 + 170612 d^2 -   255552 d + 153088) t \notag \\ 
 & -
 8 (d - 4) (8 d^6 - 15 d^5 - 1097 d^4 + 10424 d^3 - 40020 d^2 +    71440 d - 48960)
 \Bigr) \notag \\ 
+  \Shat{3} &
 \Bigl(
 (d - 4) (d - 3) (d - 2) (3 d - 8) (d^3 - 20 d^2 + 104 d - 
      176) t^4 \notag \\ 
 & +(d - 2) (21 d^6 - 405 d^5 + 3306 d^4 - 14500 d^3 +    35904 d^2 - 47584 d + 26368) t^3 \notag \\ 
 & -
 (d - 4) (177 d^6 -      3853 d^5 + 34034 d^4 - 157624 d^3 + 405240 d^2 - 548336 d \notag \\ 
 & +
 304512) t^2 \notag \\ 
 & +
 (279 d^7 - 7465 d^6 + 83720 d^5 - 514032 d^4 +   1872848 d^3 - 4053616 d^2 + 4823552 d \notag \\ 
 & -
 2430976) t \notag \\ 
 & -
   2 (63 d^7 - 1723 d^6 + 19776 d^5 - 124416 d^4 + 464784 d^3 - 
      1031536 d^2 + 1258240 d \notag \\ 
 & - 649728)
      \Bigr)
\Bigr] \notag \\
   &\hspace{-12mm}+{  \MIFiveZeroOnep  \over   16 (d - 3) (d - 2) (2 d - 7) (t - 1)^3 t (t + 1) } \notag \\ \times 
 \Bigl[
 \Shat{1} &
 \Bigl(
 4 (d - 4) (d - 2) (2 d^3 - 32 d^2 + 155 d - 234) t^5 \notag \\ 
 & - 
 4 (22 d^5 - 538 d^4 + 4547 d^3 - 17726 d^2 + 32648 d -   22912) t^4 \notag \\ 
 & +
 4 (308 d^5 - 4361 d^4 + 25693 d^3 - 79592 d^2 + 129232 d -   86272) t^3 \notag \\ 
 & +
 4 (308 d^5 - 3967 d^4 + 19561 d^3 - 45038 d^2 + 45452 d -    12496) t^2 \notag \\ 
 & -
 4 (2 d - 7) (11 d^4 - 252 d^3 + 1760 d^2 - 5004 d + 5104) t \notag \\ 
 & +
 4 (d - 4) (2 d - 7) (d^3 - 14 d^2 + 60 d - 84)
 \Bigr) \notag \\ 
+  \Shat{3} &
 \Bigl(
 (d - 3) (d - 2) (d^3 - 20 d^2 + 104 d - 176) t^6 \notag \\ 
 & -
 2 (6 d^5 - 151 d^4 + 1288 d^3 - 5082 d^2 + 9532 d - 6864) t^5 \notag \\ 
 & +
 2 (59 d^5 - 1096 d^4 + 8536 d^3 - 33882 d^2 + 66784 d -  51280) t^4 \notag \\ 
 & -
 4 (36 d^5 - 854 d^4 + 7448 d^3 - 30851 d^2 + 61538 d -   47584) t^3 \notag \\ 
 & +
 (-27 d^5 - 101 d^4 + 3466 d^3 - 18324 d^2 +  38392 d - 29216) t^2 \notag \\ 
 & +
 2 (2 d - 7) (17 d^4 - 312 d^3 + 1992 d^2 - 5424 d + 5424) t \notag \\ 
 & -
 2 (d - 4) (2 d - 7) (d^3 - 14 d^2 + 60 d - 84)
 \Bigr)
 \Bigr]  \notag \\
   &\hspace{-12mm}+{  \MIFiveZeroTwop  \over   16 (d - 4)^2 (d - 3) (d - 2) (2 d - 7) (t - 1) t (t + 1) } \notag \\ \times 
 \Bigl[
 \Shat{1} &
 \Bigl(
 4 (d - 4) (d - 2) (2 d - 9) (8 d^3 - 123 d^2 + 583 d - 870) t^3 \notag \\ 
 & -
 8 (d - 4) (2 d - 9) (52 d^4 - 546 d^3 + 2363 d^2 - 4971 d + 
    4074) t^2 \notag \\ 
 & -
 4 (d - 4) (2 d - 9) (104 d^4 - 915 d^3 + 2687 d^2 - 2654 d - 24) t \notag \\ 
 & +
 16 (d - 4) (2 d - 9) (2 d - 7) (d^3 - 14 d^2 + 60 d - 84)
 \Bigr) \notag \\ 
+  \Shat{3} &
 \Bigl(
 3 (d - 3) (d - 2) (2 d - 9) (d^3 - 20 d^2 + 104 d - 176) t^4 \notag \\ 
 & -(2 d - 
    9) (41 d^5 - 785 d^4 + 6198 d^3 - 24804 d^2 + 49192 d - 
    37888) t^3 \notag \\ 
 & +
 (2 d - 9) (57 d^5 - 1385 d^4 + 12250 d^3 - 
    51224 d^2 + 102800 d - 79712) t^2 \notag \\ 
 & -
 (2 d - 9) (3 d^5 - 331 d^4 + 
    3818 d^3 - 17188 d^2 + 34600 d - 26176) t \notag \\ 
 & -
 8 (d - 4) (2 d - 9) (2 d - 7) (d^3 - 14 d^2 + 60 d - 84)
 \Bigr)
 \Bigr]  \notag \\
   &\hspace{-12mm}+{  \text{MI601}  \over   32 (2 d - 7) } \notag \\ \times 
 \Bigl[
 \Shat{3} & t^2
 \Bigl(
 d^3 - 20 d^2 + 104 d - 176
 \Bigr)
  \Bigr]  \notag \\
   &\hspace{-12mm}+{  \text{MI601p}  \over 32 (2 d - 7)   } \notag \\ \times 
 \Bigl[
 \Shat{1} &
 \Bigl(
 4 (d - 4) (3 d - 20) - 4 (d - 4) (5 d - 22) t
 \Bigr) \notag \\ 
+  \Shat{3} & 
 \Bigl(
 d^3 - 20 d^2 + 100 d + (d^3 - 20 d^2 + 104 d - 176) t^2 \notag \\ 
 & -
    2 (d^3 - 20 d^2 + 102 d - 168) t - 160
 \Bigr)
 \Bigr] \, .
\end{align}

\section{Definition of the master integrals}
\label{sec:masterintegrals}
The two-loop master integrals are defined as 
\begin{equation}
\text{MI} =
\int
\frac{d^{d}\ell_1}{i\pi^{d/2}}
\frac{d^{d}\ell_2}{i\pi^{d/2}}
\frac{1}{\prod_{j=1}^{15} \mathcal{P}_j^{a_j}} \, ,
\end{equation}
where the propagators are given by
\begin{align}
 & \mathcal{P}_1 =\ell_{1}^{2},
 & \mathcal{P}_9    & = (p_{2} - q_{1} + \ell_{1} + \ell_{2})^{2}, \notag \\
 & \mathcal{P}_2 =\ell_{2}^{2} ,
 & \mathcal{P}_{10} & = \ell_{2}^{2}-m_{t}^{2},   \notag \\
 & \mathcal{P}_3 = ( p_{2} - \ell_{1})^{2},
 & \mathcal{P}_{11} & = ( - p_1 + \ell_{1} + \ell_{2})^{2}-m_{t}^{2},\notag \\
 & \mathcal{P}_4 = (q_{1} + \ell_{2})^{2}-m_{t}^{2},
 & \mathcal{P}_{12} & = (q_{2} - p_{1} + \ell_{1})^{2} , \notag \\
 & \mathcal{P}_5 = (\ell_{1} +\ell_{2})^{2},
 & \mathcal{P}_{13} & = (p_{2} + \ell_{1})^{2},\notag \\
 & \mathcal{P}_6 = (p_{1} - q_{2} + \ell_{1} + \ell_{2})^{2}-m_{t}^{2}, 
 & \mathcal{P}_{14} & = (q_{1} + \ell_{1})^{2} -m_{t}^{2} , \notag \\
 & \mathcal{P}_7 = (- p_{1} + \ell_{1})^{2},
 & \mathcal{P}_{15} & = (\ell_{1} - \ell_{2})^{2} -m_{t}^{2},\notag \\
 & \mathcal{P}_8 = (q_{2} + \ell_{2})^{2}, 
 &
\end{align}
with external momenta such that
$p_1+p_2 = q_1+q_2$,
$p_1^2 = p_2^2 = q_2^2 = 0$,
$(p_1-q_2)^2 = t$ and $q_1^2 = m_t^2 = 1$.
The powers of the propagators are shown for each master integral in
\Tab{tab:integrals}.
\begin{table}[ht]
\begin{tabular}{|l|c|c|c|c|c|c|c|c|c|c|c|c|c|c|c|}
\hline
   & $a_{1}$
   & $a_{2}$
   & $a_{3}$
   & $a_{4}$
   & $a_{5}$
   & $a_{6}$
   & $a_{7}$
   & $a_{8}$
   & $a_{9}$
   & $a_{10}$
   & $a_{11}$
   & $a_{12}$
   & $a_{13}$
   & $a_{14}$
   & $a_{15}$\\
\hline
\hline
MI301    & 0 & 0 & 0 & 0 & 1 & 0 & 1 & 1 & 0 & 0 & 0 & 0 & 0 & 0 & 0 \\
\hline
MI301p   & 0 & 0 & 1 & 1 & 1 & 0 & 0 & 0 & 0 & 0 & 0 & 0 & 0 & 0 & 0 \\
\hline
MI301pu1 & 1 & 0 & 0 & 1 & 1 & 0 & 0 & 0 & 0 & 0 & 0 & 0 & 0 & 0 & 0 \\
\hline
MI302p   & 0 & 0 & 2 & 1 & 1 & 0 & 0 & 0 & 0 & 0 & 0 & 0 & 0 & 0 & 0 \\
\hline  
MI303p   & 0 & 0 & 0 & 0 & 0 & 0 & 0 & 0 & 0 & 1 & 0 & 0 & 0 & 1 & 1 \\
\hline
MI401    & 0 & 1 & 0 & 0 & 1 & 0 & 1 & 0 & 1 & 0 & 0 & 0 & 0 & 0 & 0 \\
\hline
MI402p   & 0 & 1 & 1 & 0 & 1 & 1 & 0 & 0 & 0 & 0 & 0 & 0 & 0 & 0 & 0 \\
\hline
MI403p   & 1 & 1 & 0 & 1 & 0 & 1 & 0 & 0 & 0 & 0 & 0 & 0 & 0 & 0 & 0 \\
\hline
MI404p   & 1 & 2 & 0 & 1 & 0 & 1 & 0 & 0 & 0 & 0 & 0 & 0 & 0 & 0 & 0 \\
\hline
MI406t   & 1 & 0 & 0 & 0 & 0 & 0 & 0 & 0 & 0 & 1 & 1 & 1 & 0 & 0 & 0 \\
\hline
MI407p   & 0 & 0 & 0 & 0 & 0 & 0 & 0 & 0 & 0 & 1 & 0 & 0 & 1 & 1 & 1 \\
\hline
MI407t   & 2 & 0 & 0 & 0 & 0 & 0 & 0 & 0 & 0 & 1 & 1 & 1 & 0 & 0 & 0 \\
\hline
MI408p   & 0 & 0 & 0 & 0 & 0 & 0 & 0 & 0 & 0 & 1 & 0 & 0 & 2 & 1 & 1 \\
\hline
MI409p   & 1 & 0 & 0 & 1 & 1 & 1 & 0 & 0 & 0 & 0 & 0 & 0 & 0 & 0 & 0 \\
\hline
MI410p   & 1 & 0 & 0 & 1 & 2 & 1 & 0 & 0 & 0 & 0 & 0 & 0 & 0 & 0 & 0 \\
\hline
MI411p   & 2 & 0 & 0 & 1 & 1 & 1 & 0 & 0 & 0 & 0 & 0 & 0 & 0 & 0 & 0 \\
\hline
MI501p   & 0 & 1 & 1 & 1 & 1 & 1 & 0 & 0 & 0 & 0 & 0 & 0 & 0 & 0 & 0 \\
\hline
MI502p   & 0 & 2 & 1 & 1 & 1 & 1 & 0 & 0 & 0 & 0 & 0 & 0 & 0 & 0 & 0 \\
\hline
MI601    & 1 & 1 & 0 & 0 & 1 & 0 & 1 & 1 & 1 & 0 & 0 & 0 & 0 & 0 & 0 \\
\hline
MI601p   & 1 & 1 & 1 & 1 & 1 & 1 & 0 & 0 & 0 & 0 & 0 & 0 & 0 & 0 & 0 \\
\hline
\end{tabular}
\caption{Two-loop master integrals}
\label{tab:integrals}
\end{table}

In addition, the following one-loop master integrals appear:
\begin{align}
\text{MI101p} & = \int
\frac{d^{d}\ell_1}{i\pi^{d/2}}
\frac{1}{\ell_{1}^{2}-m_{t}^{2}} \\
\text{MI201} & = \int
\frac{d^{d}\ell_1}{i\pi^{d/2}}
\frac{1}{ (\ell_{1}+q_2-p_1)^2 \ell_{1}^{2}} \\
\text{MI201p} & = \int
\frac{d^{d}\ell_1}{i\pi^{d/2}}
\frac{1}{ (\ell_{1}+q_2-p_1)^2 (\ell_{1}^{2}-m_{t}^{2})} \, .
\end{align}
%

\end{document}